\documentclass[12pt]{elsarticle}

\usepackage [latin1]{inputenc}

\usepackage{lscape}
\usepackage{graphics}
\usepackage{epsfig}
\usepackage{graphicx}
\usepackage{color}
\usepackage{amsmath, amsthm} 

\begin{document}
\baselineskip=24pt

\newcommand{\D}{\displaystyle} 
\newcommand{\T}{\textstyle} 
\newcommand{\SC}{\scriptstyle} 
\newcommand{\SSC}{\scriptscriptstyle} 

\newcommand{\be}{\begin{eqnarray}}
\newcommand{\ee}{\end{eqnarray}}

\definecolor{yellow}{rgb}{0.95,0.75,0.1}
\definecolor{orange}{rgb}{0.95,0.4,0.1}
\definecolor{red}{rgb}{1,0,0}
\definecolor{green}{rgb}{0,1,0}
\definecolor{blue}{rgb}{0,0.5,1}

\definecolor{lblue}{rgb}{0,0.8,1}
\definecolor{dblue}{rgb}{0,0,1}
\definecolor{dgreen}{rgb}{0,0.65,0}
\definecolor{lila}{rgb}{0.8,0,0.8}
\definecolor{violet}{rgb}{1,0,0.9}
\definecolor{grey}{rgb}{0.3,0.3,0.3}

\definecolor{contoura}{rgb}{0,0,1}
\definecolor{contourb}{rgb}{0,1,1}
\definecolor{contourc}{rgb}{0,1,0}
\definecolor{contourd}{rgb}{0.95,0.75,0.1}
\definecolor{contoure}{rgb}{1,0,0}
\definecolor{contourf}{rgb}{1,0,1}

\newcommand\cred[1]{\textcolor{red}{#1}}
\newcommand\cblue[1]{\textcolor{blue}{#1}}
\newcommand\ccyan[1]{\textcolor{contourb}{#1}}
\newcommand\cgreen[1]{\textcolor{green}{#1}}
\newcommand\cyellow[1]{\textcolor{yellow}{#1}}
\newcommand\cviolet[1]{\textcolor{violet}{#1}}
\newcommand\cmagenta[1]{\textcolor{magenta}{#1}}
\newcommand\cgrey[1]{\textcolor{grey}{#1}}

\newcommand\cconta[1]{\textcolor{contoura}{#1}}
\newcommand\ccontb[1]{\textcolor{contourb}{#1}}
\newcommand\ccontc[1]{\textcolor{contourc}{#1}}
\newcommand\ccontd[1]{\textcolor{contourd}{#1}}
\newcommand\cconte[1]{\textcolor{contoure}{#1}}
\newcommand\ccontf[1]{\textcolor{contourf}{#1}}

\newcommand {\black} {\color{black}}
\newcommand {\violet}{\color{violet}}
\newcommand {\blue} {\color{blue}}
\newcommand {\dblue} {\color{dblue}}
\newcommand {\cyan} {\color{cyan}}
\newcommand {\lila} {\color{lila}}
\newcommand {\yellow} {\color{yellow}}
\newcommand {\green} {\color{green}}
\newcommand {\dgreen} {\color{dgreen}}
\newcommand {\magenta} {\color{magenta}}
\newcommand {\red} {\color{red}}
\newcommand {\orange} {\color{orange}}

\def\AJ{{\it Astron. J.} }
\def\ARAA{{\it Annual Rev. of Astron. \& Astrophys.} }
\def\ARNPS{{\it Annual Rev. of Nucl. \& Part. Sci. } }
\def\ApJ{{\it Astrophys. J.} }
\def\ApJL{{\it Astrophys. J. Letters} }
\def\ApJS{{\it Astrophys. J. Suppl.} }
\def\ApP{{\it Astropart. Phys.} }
\def\AA{{\it Astron. \& Astroph.} }
\def\AAR{{\it Astron. \& Astroph. Rev.} }
\def\AAL{{\it Astron. \& Astroph. Letters} }
\def\AASu{{\it Astron. \& Astroph. Suppl.} }
\def\AN{{\it Astron. Nachr.} }
\def\ASR{{\it Adv. in Space Res.} }
\def\EPJC{{\it Eur. Phys. Journ.} {\bf C} }
\def\IJMP{{\it Int. J. of Mod. Phys.} }
\def\JCAP{{\it J. of Cosmol. and Astrop. Phys.} }
\def\JGR{{\it Journ. of Geophys. Res.}}
\def\JHEP{{\it Journ. of High En. Phys.} }
\def\JPhG{{\it Journ. of Physics} {\bf G} }
\def\CQG{{\it Class. Quant. Grav. } }
\def\MNRAS{{\it Month. Not. Roy. Astr. Soc.} }
\def\Nature{{\it Nature} }
\def\NewAR{{\it New Astron. Rev.} }
\def\NewA{{ New Astron.} }
\def\NIMPA{{\it Nucl. Instr. Meth. Phys. Res.}{\bf A} }
\def\PASP{{\it Publ. Astron. Soc. Pac.}�}
\def\PhFl{{\it Phys. of Fluids} }
\def\PLB{{\it Phys. Lett.}{\bf B} }
\def\PRp{{\it Phys. Rep.} }
\def\PR{{\it Phys. Rev.} }
\def\PRC{{\it Phys. Rev.} {\bf C} }
\def\PRD{{\it Phys. Rev.} {\bf D} }
\def\PRL{{\it Phys. Rev. Letters} }
\def\PRX{{\it Phys. Rev. }{\bf X} }
\def\RMP{{\it Rev. Mod. Phys.} }
\def\RPP{{\it Rep. Pro.Phys.} }
\def\Science{{\it Science}�}
\def\SSR{{\it Space Sci. Rev.}�}
\def\ZfA{{\it Zeitschr. f{\"u}r Astrophys.} }
\def\ZfN{{\it Zeitschr. f{\"u}r Naturforsch.} }
\def\etal{{\it et al.}}

\hyphenation{mono-chro-matic  sour-ces  Wein-berg
chang-es Strah-lung dis-tri-bu-tion com-po-si-tion elec-tro-mag-ne-tic
ex-tra-galactic ap-prox-i-ma-tion nu-cle-o-syn-the-sis re-spec-tive-ly
su-per-nova su-per-novae su-per-nova-shocks con-vec-tive down-wards
es-ti-ma-ted frag-ments grav-i-ta-tion-al-ly el-e-ments me-di-um
ob-ser-va-tions tur-bul-ence sec-ond-ary in-ter-action
in-ter-stellar spall-ation ar-gu-ment de-pen-dence sig-nif-i-cant-ly
in-flu-enc-ed par-ti-cle sim-plic-i-ty nu-cle-ar smash-es iso-topes
in-ject-ed in-di-vid-u-al nor-mal-iza-tion lon-ger con-stant
sta-tion-ary sta-tion-ar-i-ty spec-trum pro-por-tion-al cos-mic
re-turn ob-ser-va-tion-al es-ti-mate switch-over grav-i-ta-tion-al
super-galactic com-po-nent com-po-nents prob-a-bly cos-mo-log-ical-ly
Kron-berg Berk-huij-sen}
\def\simle{\lower 2pt \hbox {$\buildrel < \over {\scriptstyle \sim }$}}
\def\simge{\lower 2pt \hbox {$\buildrel > \over {\scriptstyle \sim }$}}
\def\intunits{{\rm s}^{-1}\,{\rm sr}^{-1} {\rm cm}^{-2}}

\def\res{{\mathit{res}}}
\def\DIF{{\mathit{diff}}}
\def\spa{{\mathit{spal}}}
\def\INTE{{\mathit{intr}}}
\def\bub{{\mathit{bubl}}}
\def\obs{{\mathit{obs}}}
\def\pri{{\mathit{prim}}}
\def\sec{{\mathit{sec}}}
\def\sun{\hbox{$\odot$}}
\def\la{\mathrel{\mathchoice {\vcenter{\offinterlineskip\halign{\hfil
$\displaystyle##$\hfil\cr<\cr\sim\cr}}}
{\vcenter{\offinterlineskip\halign{\hfil$\textstyle##$\hfil\cr
<\cr\sim\cr}}}
{\vcenter{\offinterlineskip\halign{\hfil$\scriptstyle##$\hfil\cr
<\cr\sim\cr}}}
{\vcenter{\offinterlineskip\halign{\hfil$\scriptscriptstyle##$\hfil\cr
<\cr\sim\cr}}}}}
\def\ga{\mathrel{\mathchoice {\vcenter{\offinterlineskip\halign{\hfil
$\displaystyle##$\hfil\cr>\cr\sim\cr}}}
{\vcenter{\offinterlineskip\halign{\hfil$\textstyle##$\hfil\cr
>\cr\sim\cr}}}
{\vcenter{\offinterlineskip\halign{\hfil$\scriptstyle##$\hfil\cr
>\cr\sim\cr}}}
{\vcenter{\offinterlineskip\halign{\hfil$\scriptscriptstyle##$\hfil\cr
>\cr\sim\cr}}}}}
\def\degr{\hbox{$^\circ$}}
\def\arcmin{\hbox{$^\prime$}}
\def\arcsec{\hbox{$^{\prime\prime}$}}
\def\utw{\smash{\rlap{\lower5pt\hbox{$\sim$}}}}
\def\udtw{\smash{\rlap{\lower6pt\hbox{$\approx$}}}}
\def\fd{\hbox{$.\!\!^{\rm d}$}}
\def\fh{\hbox{$.\!\!^{\rm h}$}}
\def\fm{\hbox{$.\!\!^{\rm m}$}}
\def\fs{\hbox{$.\!\!^{\rm s}$}}
\def\fdg{\hbox{$.\!\!^\circ$}}
\def\farcm{\hbox{$.\mkern-4mu^\prime$}}
\def\farcs{\hbox{$.\!\!^{\prime\prime}$}}
\def\fp{\hbox{$.\!\!^{\scriptscriptstyle\rm p}$}}
\def\baselinestretch{1.5}
\def\gsim{\stackrel{>}{\sim}}
\def\lsim{\stackrel{<}{\sim}}
\def\beq{\begin{equation}}
\def\eeq{\end{equation}}
\def\ol{\overline}

%
\begin{centering}
{\bf \Large {Loaded layer-cake model for cosmic ray interaction around exploding super-giant stars making black holes}}
\vspace{0.5cm}\\
\end{centering}

{\bf  M.L. Allen} (Washington State University, Pullman, WA, USA); 
{\bf   P.L. Biermann} (i) MPIfR, Bonn Germany; ii) University of Alabama Tusca\-loosa, Department of Physics \& Astrononmy, USA); 
{\bf  A. Chieffi} (INAF, Rome, Italy);  
{\bf   D. Frekers} (University of M{\"u}nster, Department of Physics, M{\"u}nster, Germany); 
{\bf  L.{\'A}. Gergely} (i) Univ. Szeged, Department of Theoretical Physics, Hungary; ii) HUN-REN Wigner Research Centre for Physics, Department of Theoretical Physics, Budapest, Hungary); 
{\bf \boldmath  B. Harms$^{\dagger}$} (University of Alabama in Tuscaloosa, Department of Physics \& Astronomy, AL USA);  
{\bf I. Jaroschewski} (University of Bochum, Faculty of Physics \& Astronomy, Bochum, Germany); 
{\bf  P.S. Joshi} (International Center for Space \& Cosmology,
Ahmedabad Univ., Ahmedabad 380009, India); 
{\bf \boldmath  P.P. Kronberg$^{\dagger}$} (Univerisity of Toronto, Department of Physics \& Astronomy, Toronto, CA); 
{\bf  E. Kun} (University of Bochum, Faculty of Physics \& Astronomy, Bochum, Germany); 
{\bf  A. Meli} (i) College of Science \& Technology, Departmet of Physics
NC A\&T State University, Greensboro, USA; ii) STAR Institute, IFPA, Universite de Li{\`e}ge, Belgium);  
{\bf  E.-S. Seo}, (University of  Maryland, Department of Physics, College Park, MD, USA); 
{\bf  T. Stanev}, (Bartol Research  Institute, University of Delaware, USA) 

%
%

%
\vspace{0.5cm}

{\small {\bf \boldmath $^{\dagger}$} no longer among us}.\\
{\small Corresponding author: Peter L. Biermann, plbiermann@mpifr-bonn.mpg.de}\\
\\

{\center For the Astroparticle Physics Memorial Volume for Tom Gaisser.}

\section{Abstract}

{\small The experiment AMS on the International Space-Station has produced accurate cosmic ray spectra for many chemical elements, both primaries like He, C, O, Fe, other cosmic ray (CR) primaries like Ne, Mg, and Si, secondaries like Li, Be, B, and of mixed provenance, like N, Na, and Al. The AMS spectra demonstrate that interaction is seriously diminishing fluxes up to a rigidity of about 100 GV, and so the existing models for CR interaction have to be re-examined. Based on earlier well-established ideas a model is proposed here that focusses on the cosmic ray interaction first in the wind shock shell of super giant stars, when the supernova driven shock races through, and second in the OB-Superbubble surrounding the SN: These stars include both red super-giant stars and blue super-giant stars; both produce black holes in their explosion, and drive winds and jets with electric currents. Variability of these winds or jets gives rise to temporary electric fields, as has recently been demonstrated, and discharge (so lightning) acceleration gives steep spectra, with synchrotron losses to $p^{-5}$ in momentum $p$; these spectra are typically observed in both Galactic and some extragalactic radio filaments. Analogous hadron spectra $p^{-4}$ excite a flat spectrum of magnetic irregularities in the bubble zone, which in turn yields a steep dependence of residence time versus energy, with power $- \, 5/3$. This spectrum is indicated by the AMS data  and appears to be required to explain the CR spectra below 100 GV. The emphasis in this paper is to work out the interaction of the freshly accelerated cosmic ray particles. In the model presented here the interaction is derived as a function of time, and then integrated, or developed to long times. The model gives a rigidity dependence of the secondary/primary ratio of slope $- \, 1/3$ as well as the strong reduction of the primary fluxes below a rigidity of about 100 GV, relative to a power-law injection spectrum, with slope $+ \, 2$. The two key aspects based on blue super-giant stars and a magnetic irregularity spectrum in the bubble zone given by lightning are i) a much larger column of interaction, allowed because of heavy element enrichment of the interaction zone, and ii) even He, C, and O may have a small secondary contribution, as the difference to the Fe spectrum suggests; this small secondary component is visible in the $^3$He/$^4$He ratio. The model may also explain the spectrum of CR anti-protons, the gamma-ray spectra of the Galaxy and the high energy neutrino spectrum of our Galaxy, including also red super-giant stars as sources. ISM-SNe, i.e. SN Ia and neutron-star SNe, contribute to CR protons and CR He nuclei.}

\section{Introduction}

The origin of the cosmic-ray (CR) particles has been traced to supernova (SN) explosions since 1934 \cite{Baade34}. In the 1980s \cite{Prantzos84,Voelk88} models started to emphasize that very massive stars all have winds, that influence the CR particle injection, and acceleration \cite{Drury83,Hillas84,Prantzos84,Prantzos86,Prantzos91,Prantzos93,Prantzos95,Prantzos11,Prantzos12a,Prantzos12b}, and so also their chemical composition (e.g., \cite{CR-I,CR-II,CR-III,CR-IV,Biermann95,Ellison97,Wiebel98,Nolfo06,Strong07,Thoudam16,Gupta18,Gupta20}).

There are clearly several source classes of CR particles: Supernovae, that explode into the Interstellar Medium (ISM), ISM-SNe, can be of two kinds: exploding white dwarfs (SN Ia), and massive stars in the mass range about 10 $M_{\odot}$ to 25 $M_{\odot}$ Zero Age Main Sequence (ZAMS) mass that make neutron stars in their explosion \cite{Heger03,Chieffi13,Limongi18}. Another class of sources is massive stars above this mass range that explode into their prior wind, wind-SNe. These stars produce a Red Super Giant (RSG) before the explosion, for about 25 to 33 $M_{\odot}$ ZAMS mass or a Blue Super Giant (BSG) star for most masses above 33 $M_{\odot}$ ZAMS mass before the explosion (\cite{Chieffi13}, summarized with uncertainties and rates in \cite{ASR18}). These mass ranges strongly depend on abundances, here given for cosmic abundances, and high rotation rates. At some much higher mass there is the possibility that there is no remnant at all \cite{Heger03}. The RSG and BSG star explosions usually produce a black hole (BH) as their remnant. These stars all make black holes up to about $80 \, M_{\odot}$ ZAMS mass; above that stars either blow up completely, make a neutron star, or a black hole \cite{Heger03,Limongi18}. These black holes may spin quite fast initially, but lose their rotational energy rapidly through magnetic winds \cite{BZ77,Limongi18}. Analogously neutron stars can spin very rapidly and inject a lot of energy into their surroundings from spin-down. The exact transition ZAMS mass between neutron stars and black holes depends on heavy element abundance, and rotation (see also \cite{Mirabel11}).  RSG stars have a slow dense wind, and BSG stars have a fast tenuous wind. A wind shock shell accumulates the most recent material ejected from the star, and therefore gets heavily enriched in the case of BSG stars \cite{Sina01}. That wind shocked shell is embedded in a wind-SN-superbubble, the remnants of the giant molecular cloud, or other interstellar material, out of which the massive stars were born. Therefore a SN shock traversing this wind can be expected to yield very different properties for the SN-driven shock running through old RSG and old BSG star winds. And yet the observations of RSG and BSG star explosions \cite{ASR18,Gal19} show that shock speed, prior mass loss, and the magnetic field (in terms of $B \, \times \, r$) are consistent with the same numbers, for the quantity $B \, \times \, r \, = \, 10^{16.0 \pm 0.12} \, {\rm Gauss \, cm}$; here $B$ is the magnetic field strength observed at distance $r$ from the center. This specific number is based on the radio-supernovae (RSNe) in the starburst galaxy M82 \cite{KBS85,Allen98,Allen99}. There are other possible sources of energetic particles, like pulsar-wind nebulae, stars orbiting a black hole or neutron  star, and of course active galactic nuclei (AGN). The recent AMS data on energetic particles \cite{AMS18,AMS19a,AMS19b,AMS19c,AMS20,AMS21a,AMS21b,AMS21c,AMS21d,AMS21e,AMS21f} challenge all these sites for CR origin. In this paper the focus is mostly on the question of what RSG and BSG star explosions can contribute to the CR particles.

Here the first focus is on blue super giant stars (BSG), which can get heavily enriched in their wind shock shell. When a super-giant star explodes \cite{Woosley02}, a shock wave is sent through its wind, loaded with energetic (CR) particles, possibly with some extra driving from their central compact object, a black hole \cite{Goldreich69,Pacini73,BZ77}.  However, the approach can also be used for RSG stars, which are usually not enriched by much. For them spallation would be a much weaker process due to the low enrichment.

Comparing the data for the larger RSNe in M82 with those observed immediately after the initial explosion allows to cover a range of radii of more than a factor of 100 \cite{Gal19}, and yet the quantity $B \, \times \, r$ remains the same, so matching what is expected on the basis of a Parker wind \cite{Parker58} with a constant velocity.

Tracing the origin of CR particles from source to detector involves transit through first the source region, then the interstellar medium (ISM) full of dense clouds, of order 100 pc thick, and also the tenuous phase of the ISM, which is of order kpc thick. Leakage from the ensemble of clouds and tenuous medium full of Kolmogorov turbulence \cite{Biermann15,Gaisser13} produces a spectrum steeper by 1/3 as compared to the sources. Therefore it is natural to assume that also most interaction happens in this ensemble of clouds and tenuous medium \cite{GarciaMunoz87,Simpson88}. However, already early it was noticed that most massive stars agglomerate in associations, forming an OB-Superbubble \cite{Higdon98,Higdon05,Binns05,Lingenfelter07,Binns11b,Higdon13}, full of HII-regions, OB stars, the effect of the powerful winds of such stars, and finally also the remnants of massive stars, neutron stars and black holes. Neutron stars and black holes eject winds and jets, relativistic in the case of black holes. The dense environment of massive star winds, highly enriched in heavy elements, easily matches the column density required for interaction by the data \cite{Biermann01,Sina01}.

One question has always been what is the maximum particle energy that can be reached by charged particles in the supernova shock \cite{Cox72,Lagage83,Hillas84,Jokipii87,Voelk88,CR-I,Lucek00,Bell01}, and so whether the ankle energy can be reached for Galactic sources. The direct observations (reviewed in \cite{ASR18,Gal19}, with the magnetic field numbers mentioned above) show that this is possible, noting that the magnetic fields in winds are tangential \cite{Parker58,Weber67}, and so Jokipii's arguments apply about perpendicular shocks \cite{Jokipii87}. Using the spatial limit criterion \cite{Hillas84} also demonstrates that this is doable. The observed post-shock magnetic field \cite{ASR18,Gal19} allows particles of charge $Z$ to reach the ankle energy for a range of nuclei. This gives a maximum particle energy of $Z \, 10^{17.6 \pm 0.12}$ eV. Therefore all such supernovae (in our Galaxy averaging every 600 years for BSG stars, and every 400 years for both RSG and BSG stars together, \cite{Diehl10}) accelerate particles to EeV energies, so could be called Exatrons, a factor of $10^3$ beyond PeV sources or Pevatrons. 

Now with the highly accurate spectra that AMS had obtained \cite{AMS18,AMS19a,AMS19b,AMS19c,AMS20,AMS21a,AMS21b,AMS21c,AMS21d,AMS21e,AMS21f}, as well as ISS-CREAM data \cite{ISS21a,ISS21b,ISS22}, the challenge to understand the CR spectra has intensified, and this paper is a proposal of an explanation of the new AMS CR spectra together with $\gamma$-ray \cite{FermiLAT11}, neutrino data \cite{IceCube23}, massive star statistics \cite{Chini12,Chini13a,Chini13b} as well as $\gamma$-ray line statistics \cite{Prantzos95,Diehl10,Prantzos11,Prantzos12a,Prantzos12b}. One test will be whether the model can also help understand Galactic gammas, neutrinos, positrons and anti-protons. However, the model worked out in detail is built on the ideas of SN-explosions in OB-Superbubbles \cite{Castor75}, and so is founded in ideas that were established decades ago, and developed over some time \cite{Higdon98,Higdon05,Binns05,Lingenfelter07,Binns11b,Higdon13}. The model proposed is shown as a series of mathematical expressions, that remain to be properly fitted to the data. Finally, does this model make any predictions, that could allow one to rule it out?

\subsection{Ubiquitous lightning}

Recently it was noted \cite{Gopal24} that winds and jets carry electric currents \cite{Parker58,Weber67}. In the case that such winds or jets vary with time, the electric currents also inevitably vary with time. When these currents vary rapidly with time they build up electric charges, and so electric fields. These electric fields discharge, and in the discharge accelerate particles again. In the limit that electric fields dominate over magnetic fields in the scattering the spectrum is expected to be a 1D equipartition spectrum, so $p^{-2}$ in 1D in momentum $p$, which upon scattering turns into $p^{-4}$ in 3D. With losses \cite{Kardashev62} this spectrum can steepen to $p^{-5}$ for electrons. When magnetic fields dominate in scattering the limit is $p^{-2}$ in 3D, and with losses $p^{-3}$. For radio synchrotron emission this implies an expected range of spectra between ${\nu}^{-2}$ and ${\nu}^{-1}$, which is just the range observed in both Galactic and extragalactic radio filaments \cite{Zadeh22,Ramatsoku20,Dabbech22}, and so is consistent with the model proposed in \cite{Gopal24}. Lightning is strongly space and time-dependent, and so excites waves maximally \cite{Drury83}. Considering the relationship between CR spectrum $p^{-x}$ and the turbulent spectrum $I(k) \, \sim \, k^{-\beta}$ in \cite{Drury83,Biermann01}  in wave-particle spectrum interaction the following relationship is obtained

\begin{equation}
- \, \frac{1}{2}  \, - \frac{3}{2} \, \beta \; = \; x \, - 5 \, .
\end{equation}

Here the left hand side derives from the turbulent cascade equation, and the right hand side from the excitation equation \cite{Biermann01}. Hence a proton spectrum of $p^{-4}$ excites a spectrum of magnetic irregularities of $I(k) \, \sim \, k^{-1/3}$, where $I(k) \, k$ is energy density, with the rigidity ${\tilde{R}} \, \sim \, 1/k$. This results in a diffusion time dependence of ${\tilde{R}}^{- 5/3}$ in the case of lots of lightning. Obviously such a strong energy dependence hits a limit, when the diffusion time corresponds to the light travel time, or possibly even earlier, when it crosses a flatter spectrum of magnetic irregularities. The data may allow us to test such transitions.

The many radio filaments described by \cite{Zadeh22} can be interpreted as emanating from recently formed stellar mass black holes \cite{Gopal24}; they have the spectra expected, and are consistent with the lightning model also in synchrotron age, statistics, multiplicity, and location.

The lightning model \cite{Gopal24} requires a population of particles with a $E^{-2}$ spectrum in the source, close to black holes, or a spectrum very close to this. One may ask, whether such a population is visible in the CR data: it is \cite{Gaisser13}, where pop 3 has the correct properties to be interpreted as such: In the source this population has a spectrum very close to $E^{-2}$ (see below) and the upper limit energy given by $(B \times r) \, = \, 10^{16.0 \pm 0.12} \, {\rm Gauss \times cm}$ \cite{ASR18,Gal19,Auger20}. This component can be attributed to recently formed young stellar mass BHs, still rotating quite rapidly \cite{Chieffi13,Limongi18,Limongi20}.

A region full of neutron stars and black holes clearly has variable winds and jets, and those produce lightning, which gives an additional steep CR particle spectrum and the resulting spectrum of magnetic irregularities.

\subsection{Source to observer}

In the model proposed we work out the spectrum of the CR particles, as they progress from their original source spectrum through interaction and bubble zones to the observer, moving through the Interstellar Medium, which has a Kolmogorov spectrum of magnetic irregularities \cite{Gaisser13,Biermann15}, which adds a slope of $ - \, 1/3$. The source model which we use was proposed earlier \cite{CR-I,CR-II,CR-III,CR-IV}, and yields for the wind-SN-CR-component an initial slope of $- \, 7/3 \, - \, 0.02 \, \pm \, 0.02$ below the knee energy, and a $- \, 2.74 \, - \, 0.07 \, \pm  \, 0.07$ above the knee, and a slightly steeper ISM-SN-CR component $- \, 2.42 \, \pm \, 0.04$ \cite{Biermann95SSR}.

For the wind-SNe there is also a flatter polar cap component, that is $1/3$ flatter at the source, so $E^{-2}$ at the source, and all heavy nuclei detected by AMS show such an expected upturn \cite{AMS21a,AMS21b,AMS21c,AMS21d,AMS21e,AMS21f} at high energy. Especially visible in the AMS-data are the upturns in CR He, CR C, and CR O. The model here \cite{CR-I,CR-II,CR-III,CR-IV} predicts that all primary CR components show this upturn.

Thus, these particles have to move through the hot interstellar medium, with a scale height of order 1 - 2 kpc, as well as ubiquitous molecular clouds in a much thinner layer. This transit has been worked out in a model to explain the Fermi-$\gamma$-ray data  \cite{deBoer17} of the Galactic disk; in that model both the regions around the sources, and the molecular clouds encountered while in transit were separately included. Appenzeller \cite{Appenzeller74} has shown how magnetic field lines permeate clouds, which implies that energetic particles can indeed penetrate them.

The spectra which we observe approach at high energy the source spectra steepened by the Kolmogorov diffusion losses \cite{Gaisser13,Biermann15}, so steeper by $1/3$ as compared to the source spectra. As noted spallation losses are independent of energy to a good approximation at high energy.  Therefore we compare the high energy extrapolation to low energies with the actual spectra observed, so as to isolate the effects which we attribute to what happens close to the source, in our model proposal the interaction zone and the bubble zone.

\subsection{Scarcity of sources}

The model uses OB-Superbubbles, areas which are full of OB-stars, and so skew any statistics, that rely on an even and homogenous distribution of stars both in space and time. The $\gamma$-ray line data \cite{Prantzos95,Diehl10,Prantzos11,Prantzos12a,Prantzos12b} show that RSG stars blow up at a rate of one every 400 years across the Galaxy, and BSG stars one every 600 years (also summarized in detail in \cite{ASR18}). Assuming that all sources within 1 kpc radius contribute to our CR particle population the star formation in the Solar neighborhood is of order 0.01 the sum of the Galaxy \cite{Smith78}: This translates to a time scale of BSG stars exploding, for instance, of one every 60,000 years. BSG stars are, however, not all the same, but need to be differentiated as sources for various abundances (see Table 7 in \cite{Chieffi13} for how ejected abundances change with initial mass for $^4$He, $^{12}$C, $^{16}$O, and $^{56}$Fe, and other important elements and isotopes); this suggests that we need to differentiate the BSG stars into several sub-groups, at least 3 groups. This in turn implies a time-scale of 200,000 years. The time-scale to diffuse out of the Galaxy is of order $10^7$ years at GeV energies for protons \cite{Binns16}, and so $10^{5.7}$ years at 10 TeV/nucleon, using again a Kolmogorov spectrum of magnetic irregularities \cite{Biermann15}. This translates into about 10 sources contributing, or a variation of about $1/\sqrt{10}$ in CR flux due to scarcity of sources, so a variation of order 30 percent is possible at 10 TV rigidity. Correspondingly, spectral wiggles in heavy elements beyond 30 TeV would have to be interpreted with great uncertainty.  One consequence is that we might need to proceed with extreme caution exploring contradictory consequences from slightly different spectra of different elements.

However, we need to account for the fairly even spectral behavior across all energies to EeV energies \cite{Gaisser13,Auger20}; that means that even there we require at least 3, perhaps 30 sources to contribute. In the framework enumerated above this means that for, say, ten sources at EeV we need order $10^4$ sources at GeV. This means for an area of, say, 2 kpc radius $0.04 \, r_{0.3}^2$ of all BSG stars, so $15,000 \, r_{0.3}^{- \, 2}$ years for every BSG star, or $10^{8.2}\, r_{0.3}^{- \, 2}$ years. Here $r_{0.3}$ is the radius of the region considered in units of 2 kpc. This entails that the area has to be much larger, covering much of the Galaxy. This condition implies that beyond the hot disk carrying the CR particles far and wide, we may also need scattering via the initial halo wind of the Galaxy, with another scattering law than Kolmogorov. This is in fact the conclusion in \cite{Biermann15}, where it was proposed that at larger scales first a saturated law takes over, $k^{-1}$ instead of Kolmogorov $k^{-5/3}$, and at yet larger scales a shock dominated law, with $k^{-2}$, so an energy-independent scattering law. However, that still implies that at large energies we ``see" many of the BSG explosions in the disk; the analysis of \cite{Gaisser13} demonstrates that the Kolmogorov law applies to at least EeV. Since in the Solar neighborhood the star formation rate per ${\rm kpc}^2$ (Table 4, column 5 in \cite{Smith78}) is a minimum, and within 3 kpc already 10 times the rate around the Sun, all we need to do is cover 3 kpc instead of 2 kpc to reach $10^{7.2}\, r_{0.5}^{- \, 2}$ years. This is a quite possible extrapolation. However, if the real number of contributing BSG star explosion sources were smaller, that could introduce small wiggles in the spectrum observed, especially for high $Z$ elements.

\subsection{Interaction depth limits}

In CR particle propagation the data are used to derive the amount of matter traversed during propagation \cite{Binns16}. However, here those particles are not included in the analysis, that never make it to Earth. It is assumed that of all specific particles a sufficient number make it through to Earth, so that they can be counted, and compared with other particles that interact less, but derive from the same sources. 

It has been shown that the interaction depth of BSG star wind shock regions is sufficient to explain the column derived from the CR data \cite{Sina01}. In fact, in the standard derivation it is normally assumed, that the chemical composition of the matter traverse is average; however, as noted \cite{Sina01}, BSG stars have winds which are heavily enriched, the wind shock shell is also enriched, so the real interaction depth is larger than the normally quoted numbers would suggest.

This can be tested via neutrino emission, comparing models such as \cite{deBoer17,Breuhaus22} with observation \cite{IceCube23}.

Using the average magnetic field strength of $5 \, {\rm \mu Gauss}$ \cite{Beck96} we obtain an energy density of $10^{-12.0} \, {\rm erg/cm^3}$, matching an old and well established estimate. Our Galaxy has a wind, and so the Alfv{\'e}n velocity has to match the escape speed giving an upper limit to the density at the base of the halo. The escape speed can be estimated by noting that far out the mass of the Galaxy is at least $2 \, 10^{12} \, M_{\odot}$ \cite{Kahn59,Sofue01,Gaia18}, at a scale of about 200 kpc, giving an escape speed of $10^{7.3} \, {\rm cm/s}$. The Alfv{\'e}n velocity either match this or better exceed this to make an escape possible, giving an upper limit to the density at the base of $n \, < \, 10^{-2.6} \, {\rm cm^{-3}}$. With a residence time estimate of $10^7 \, {\rm yrs}$ this gives a grammage of about $10^{- \, 1.3} \, {\rm g/cm^2}$. In addition the Alfv{\'e}n velocity provides an upper limit to CR propagation. On the other hand we need to be able to reach several kpc in order to have a sufficient number of BSG explosions, as argued above, and so we obtain a second estimate of the Alfv{\'e}n velocity from reaching several kpc in $10^7 \, {\rm yrs}$, which gives an even slightly higher numerical estimate of $10^{7.5} \, {\rm cm/s}$.  In order to reach the required number of BSG star explosions, the transport of CR has to go through this hot medium of low density.  Therefore, as argued in \cite{Sina01} all CR interaction to explain secondaries, including the radioactive isotopes, has to happen close to the sources. For the CR population observed there is only negligible interaction along the path from just outside the sources to us.

\section{The layer cake model}

The model proposed here initially centers on the fully accelerated CR particle spectrum carried by the SN shock, as it hits the enriched shocked wind shell. In a second step it focusses on the propagation and interaction through the OB super-bubble. And finally these particles escape to the outside, to be considered as the source distribution as seen from Earth.

We estimate how much time it takes to inject the CR particles from the SN shock shell into the wind-shock region, we note that the typical size of the wind is a few parsec, rarely of order 10 pc (see the numbers for M82 \cite{KBS85,Allen98}). Comparing the sums in column 5 and 6 in table 2 of \cite{ASR18} gives a shock speed $U_{shock}$ of about $0.16 \, c$; for an adiabatic gas constant of $5/3$ and assuming a strong shock the flow behind the shocks runs at $(3/4) \, U_{shock}$, so $0.12 \, c$, with a relatively large range of scatter. A shock region in a wind has the thickness relative to current radius $r$ of $r/4$ \cite{CR-I,CR-II}. There is no tendency of the shock to slow down with time. Therefore the data show that the flow behind the SN-shock runs at about $0.1 \, c$, based on many observed RSNe. At 10 pc that gives a time scale of 80 years. This is quite short, and so the injection itself is taken to be instantaneous relative to the next escape time scale. This is one important step in the model proposed, that this time scale is  very much shorter than the diffusion times out of the interaction zone. That time is turn is assumed to be very much shorter than the diffusion time out of the bubble zone.

The {\bf \boldmath total number of primary particles $N_p$}, as a function of time $t$ and rigidity $\tilde{R}$, initially $N_{p, 0}$, is considered, and secondary $N_s$, within an interval of rigidity $d \, \tilde{R}$.  The key model assumption is that throughout the set of equations the calculation sticks to a rigidity and an interval. Since the nucleon number $A$ is approximately twice the charge number $Z$ for relativistic energies rigidity $\tilde{R} \, = \, (p \, c)/(Z \, e)$ (where momentum $p$ and charge $Z$ combine) can be replaced in the functional relationships for nuclei by approximately the energy $E$ per half the nucleon number $A$. Spallation,  i.e. breakup into secondary particles with number $N_s$, reduces energy and momentum, as well as charge $Z$ by about the same factor; dealing with different isotopes of the same elements, such as Fe, introduces correction factors, which can be derived from the data. If the spallation products are unstable nuclei, the framework adopted here would need to be expanded. Only in the case that the spectrum was to be very curved, or with a secondary very broad in their own distribution over rigidity, could this procedure yield corrections which themselves depend on energy.  The energy dependence is only explicitly used in some of the arguments, for historical clarity. All time scales used like ${\tau_{\INTE, \DIF, p}}$, ${\tau_{\INTE, \spa, p}} $, ${\tau_{\INTE, \spa, s}}$, etc. have a dependence on rigidity $\tilde{R}$. ${\tau_{\INTE, \DIF, p}}$ is the time-scale for diffusion from the interaction zone into the next outer region, ${\tau_{\INTE, \spa, p}} $ is the time-scale for spallation of primaries in the interaction zone, ${\tau_{\INTE, \spa, s}}$ is the time-scale for spallation of secondary nuclei in that zone, and ${\tau_{\INTE, \DIF, s}}$ is the time-scale for diffusion of secondaries to escape again from that zone, assumed to be generally the same function as for primaries. The corresponding time-scales for the bubble zone have a similar notation, but of course replace $\INTE$ with $\bub$.  The AMS data have such a precision over several decades of rigidity, that even weak energy dependencies of the model can be tested.

This can be approximated by the following equation

\begin{equation}
\frac{d \, N_{\INTE, p}}{d \, t} \; = \; - \, \frac{N_{\INTE, p}}{\tau_{\INTE, \DIF, p}} \; - \, \frac{N_{\INTE, p}}{\tau_{\INTE, \spa, p}} \; 
\end{equation}


\noindent for the primary CR particle population number $N_p$, and the time-scales $\tau_{\INTE, \DIF, p}$ for diffusion out of the interaction region, and $\tau_{\INTE, \spa, p}$ for loss of particles due to spallation. 

The SN-shock carrying a load of accelerated CR particles injecting them into the wind-shock region defines the initial condition. This initial condition is the full number of primary particles injected $N_{p, 0}$, a spectrum of energy or rigidity  $ \tilde{R}$.  For the secondary particles the equation is correspondingly

\begin{equation}
\frac{d \, N_{\INTE, s}}{d \, t} \; = \; + \, \frac{N_{\INTE, p}}{\tau_{\INTE, \spa, p}} \, - \, \frac{N_{\INTE, s}}{\tau_{\INTE, \DIF, s}} \; - \, \frac{N_{\INTE, s}}{\tau_{\INTE, \spa, s}} \; 
\end{equation}

Obviously, since baryon number is conserved, there is no spallation for proton-proton or proton-photon (gamma) interactions, but pion or meson production losses can replace the spallation loss term. If the remaining protons have a large energy loss \cite{Kelner06}, they move out of the rigidity bin used for the calculation, and so represent a total loss for the numbers of protons in that bin.

The approximation here is to ignore significant energy losses. This means in the language introduced here, that particles move to a different bin of rigidity. Considering losses, say, for electrons or positrons,  an energy bin could be used that is comoving with the losses in energy space. This is a useful approximation for adiabatic losses, synchrotron and inverse-Compton (IC) losses, dealt with separately, but difficult for the Klein-Nishina regime of IC losses. However, only one kind of loss could be considered at the same time because the energy dependence of the losses determines the comovement in energy space. For nuclei or protons the maximal energy possibly given by losses could be critical (e.g. \cite{Biermann87,Stanev10,Gaisser16}), which defines the range of validity of the approximation used here.

The procedure followed is then to derive the time-dependent behavior of all species considered, and then either integrate over time, or let time go to infinity to obtain what is expected to be observed. Everywhere, all terms are functions of rigidity $\tilde{R}$, within an interval of rigidity $d \, \tilde{R}$. As a didactic example the focus is on some unspecified primary particles (index $p$) and some secondary particles (index $s$).

Secondary spallation may be small, but as can be seen below, it may be important for heavy secondaries. As \cite{Sina01} have shown, the large mass loss in blue super giant stars prior to the supernova explosions, with heavy enrichment, implies that the wind-shock region could be strongly enriched relative to the ISM. Therefore the interaction loss from spallation might be quite large, and this is borne out by the AMS data.

For the interaction zone the combined time scales $\tau_{\INTE, \star, p}$ and $\tau_{\INTE, \star, s}$ are defined by

\begin{equation}
\frac{1}{\tau_{\INTE, \star, p}} \; = \; \frac{1}{\tau_{\INTE, \DIF, p}} \, + \, \frac{1}{\tau_{\INTE, \spa, p}} \,\,\,\,\, {\rm and} \,\,\,\,\,
\frac{1}{\tau_{\INTE, \star, s}} \; = \; \frac{1}{\tau_{\INTE, \DIF, s}} \, + \, \frac{1}{\tau_{\INTE, \spa,  s}} \, 
\end{equation}

and correspondingly for the bubble zone

\begin{equation}
\frac{1}{\tau_{\bub, \star, p}} \; = \; \frac{1}{\tau_{\bub, \DIF, p}} \, + \, \frac{1}{\tau_{\bub, \spa, p}} \,\,\,\,\, {\rm and} \,\,\,\,\,
\frac{1}{\tau_{\bub, \star, s}} \; = \; \frac{1}{\tau_{\bub, \DIF, s}} \, + \, \frac{1}{\tau_{\bub, \spa,  s}} \, 
\end{equation}

\vspace{0.3cm}

It is assumed generally that the time scales in the bubble zone are all significantly larger than in the interaction zone. That is important to simplify the mathematical expressions.

Note that the diffusion time scale usually has an energy dependence, while the spallation has a threshold, but only a weak energy dependence \cite{Ferrari14}, which will be ignored for clarity in the following. Usually the diffusion time scale decreases with energy (or rigidity), so that for large energies (and so short time scales) ${\tau_{\INTE, \star, p}}$ or ${\tau_{\INTE, \star, s}} $ approach the diffusion time scale, while for low energies (so long time scales), they approach the spallation time scale. Here the diffusion time scale is used to describe the escape from one zone to the next, or to the outside, the ISM.

Here it is not specified what secondary element/isotope nucleus is considered; but one could think of an iron (Fe) nucleus as an example of a primary, and a Lithium (Li) nucleus as a secondary. Further destruction of a secondary nucleus to another secondary (or tertiary) nucleus is allowed.

Next, the possible energy dependencies of diffusion should be discussed: There are several regimes possible, each resulting in a different momentum/charge dependence; the data will tell the observer what regime, if any here identified, is valid in nature. The spectrum of magnetic irregularities can be written as $I(k)$ energy per wavenumber $k$ and volume, and isotropy is assumed except for Kraichnan \cite{Kraichnan65}. In such a language, for instance, Kolmogorov implies $I(k) \, \sim \, k^{-5/3}$. Obviously, by resonance $1/k \, \sim \, \tilde{R} \, \sim \, E/Z$ for relativistic energies. 

The different regimes are:

(1) Saturated turbulence \cite{Goldstein95}, where the energy of the irregularities is maximal and the same per log bin of wave-number, $I(k) \, \sim \, k^{- \, 1}$.

(2) An irregularity spectrum induced by a CR spectrum \cite{Drury83} by wave-particle interaction of, e.g., $E^{- \, 7/3}$, giving $I(k) \, \sim \, k^{- \, 13/9}$. In \cite{ASR18} the general expression is discussed, as in \cite{Biermann01}.

(3) Kolmogorov, $I(k) \, \sim \, k^{- \, 5/3}$ \cite{Kolmogorov41a,Kolmogorov41b}. The Kolmogorov spectrum can be understood both as induced by a CR particle spectrum of $E^{-2}$, and by a turbulent cascade from a largest injection scale (see e.g. \cite{Biermann01}).

(4) Kraichnan \cite{Kraichnan65}, with limited dimensionality of the irregularities, $I(k) \, \sim \, k^{- \, 3/2}$.

(5) Convection limit, when convective motions give a scattering independent of energy/momentum. This gives the same spectrum  as shock-induced sawtooth irregularities, $I(k) \, \sim \, k^{- \, 2}$. Such a limit also holds if specific scales dominate the transport, as may happen in highly oblique shocks \cite{Jokipii87,CR-I,CR-II,CR-III,CR-IV,Biermann95}, like the shock thickness scale itself. Highly oblique shocks are the norm in Parker-limit winds \cite{Parker58}, where the magnetic field component $B_{\phi} \, \sim \, 1/r$, and so dominates over radial magnetic fields $B_{r} \, \sim \, 1/r^2$; the component $B_{\theta}$ is assumed to be negligible in the Parker limit \cite{Parker58}, but would be also $B_{\theta} \, \sim \, 1/r$. So allowing both $B_{\phi}$ and $B_{\theta}$ leads to perpendicular shocks for the magnetic field configuration of a SN-shock.

(6) Lightning limit: $I(k) \, \sim \, k^{- \, 1/3}$.

The scattering coefficient is then given by (e.g. \cite{Drury83})

\begin{equation}
\kappa \; = \; \frac{4}{3 \, \pi} \, r_g \, v \, \frac{B^2}{8 \, \pi \, I(k) \, k} 
\end{equation}

\noindent where $r_g$ is the Larmor radius of a gyrating charged particle, and $v$ the velocity of the particle, usually close to the speed of light $c$. The Larmor radius $r_g$ goes as $ \tilde{R}$.



So in the six cases listed above this leads to a time scale on momentum $p$ or energy $E$ dependence (for relativistic particles $E \ = \, p \, c$) of $E^{\alpha}$, or rigidity $\tilde{R}^{\alpha}$:

(1) Saturated turbulence :  $\tau_{\DIF} \; \sim \, \tilde{R}^{-1}$

(2) Induced by, e.g., a CR spectrum of  $\tilde{R}^{-7/3}$: $\tau_{\DIF} \; \sim \, \tilde{R}^{-5/9}$.

(3) Kolmogorov \cite{Kolmogorov41a,Kolmogorov41b}: $\tau_{\DIF} \; \sim \, \tilde{R}^{-1/3}$

(4) Kraichnan: $\tau_{\DIF} \; \sim \, \tilde{R}^{-1/2}$

(5) Convection or oblique shock limit: $\tau_{\DIF} \; \sim \, 1$.

(6) Lightning limit: $\tau_{\DIF} \; \sim \, \tilde{R}^{-5/3}$

One can argue that in nature all six regimes might be relevant; the data suggest certain regimes are preferred (1), (2), (3), (5), as discussed in \cite{ASR18}, and (6) as discussed in \cite{Gopal24}.  The Kolmogorov regime (3), the convection limit (5), and the lightning limit (6) are probably the most relevant.

Since the spallation time scale can be expected to be mostly independent of energy, above some threshold, it requires that the combined timescale $\tau_{\star}$ moves from a constant $\tau_{spall}$ at lower energies to an energy dependence at higher energy $\tau_{diff}$. The ratio $\tau_{\star}/\tau_{\DIF}$, e.g., then moves  from an inverted dependence of $\tau_{\spa}/\tau_{\DIF}$ at lower energies and long time scales, to unity at high energies with short time scales. The ratio $\tau_{\star}/\tau_{\spa}$, e.g., moves from unity at low energies, with long time scales, to $\tau_{\DIF}/\tau_{\spa}$ for high energies, or short time scales. For didactic clarity we left out the other indices here.

 The original equation for $N_p$ can be written as

\begin{equation}
\exp{\{- \, t/\tau_{\INTE, \star, p}\}} \, \frac{d }{d \, t} \, {\left(\, \exp{\{+ \, t/\tau_{\INTE, \star, p}\}} \,N_{\INTE, p} \, \right)} \, = \, 0 . 
\end{equation}

The solution for the primary particles then is

\begin{equation}
N_{\INTE, p} \; = \; N_{p, 0} \,  \exp{\{- \, t/\tau_{\INTE, \star, p}\}} \, 
\end{equation}

Here it needs to be emphasized that all acceleration has already happened before this point zero in time, and $N_{p,0}$ corresponds to the input spectrum of the primary particles considered, so presumably a power-law in momentum \cite{Drury83} possibly up to the maximum energy allowed by the magnetic field and the space available \cite{Hillas84}. When drift acceleration arises in addition to diffusive shock acceleration, the resulting spectrum might have a kink \cite{CR-IV}. Diffusive loss from the disk of the Galaxy steepen this spectrum, probably by the Kolmogorov turbulent cascade \cite{Berezinsky90,Gaisser13,Biermann15}.  Here the emphasis is on the difference to the source spectrum in most subsequent discussion. Using the AMS fits at face value implies that the source spectrum is close to $E^{- \, 2.38}$.  However, it will be noted below that there may be an extra component from spallation, that changes the real source spectrum, for He, C, and O, flattening it by a small amount; spallation can also lower a CR spectrum significantly. The only property they have in common is that at their center sits a black hole, possibly initially rotating fast \cite{Limongi18}.

In our case, when we derive the time-dependent interaction of energetic particles, we need to sum up all production over the time available. Since the spectrum of energetic particles is time-dependent we require an integral over time.  

To set an example the production of anti-protons can be described by

\begin{equation}
\frac{d \, n_{\bar{p}}}{d \, t} \, = \, n_{\pri} \, {\sigma_{\pri -> \bar{p}}} \, c \, n_{matter}
\end{equation}

\noindent where $n_{\bar{p}}$ is the number density of anti-protons, $\sigma_{\pri -> \bar{p}}$ is the production cross-section for anti-protons from collisions of primary hadrons of density $n_{\pri}$ with matter of density $n_{matter}$. Just in this paragraph we use $\pri$ for the primary particles, while elsewhere we use $p$ only, due to the possibility of confusion with protons versus anti-protons. This expression can be used at any rigidity separately; for a more detailed consideration the right hand side would have to be an integral over the production distribution from all rigidities of the interaction of the primary particle to some specific rigidity of the anti-proton produced. Also we would need an integral over all primary particles and matter targets that produce an anti-proton at the rigidity desired.  Then we integrate over space to get from $n_{\bar{p}}$ to $N_{\bar{p}}$, the total number of anti-protons, and from $n_{\pri}$ to $N_{\pri}$, the total number of primary cosmic ray particles. For the integral it is important, that we distinguish regions of different density, like in the interaction zone, and the bubble zone, and therefore we distinguish $N_{\INTE}$ from $N_{\bub}$. All terms are independent of time except the primary number of particles $N_{\pri}$, and so in consequence the number of, in this example of anti-protons, $N_{\bar{p}}$. To obtain the final and total number of anti-protons produced requires therefore the time-integral of $N_{\pri}$. Since the time of this integration is relatively short compared to other time-scales outside the source region the total number is important. If the further interaction and loss of the anti-protons -- or neutrinos and $\gamma$s -- is estimated to be sufficiently important to modify their spectrum, additional considerations are required.

Such an equation, with all caveats stated, can be written as

\begin{equation}
\frac{d \, N_{\bar{p}}}{d \, t} \, = \, \frac{N_{\pri}}{\tau_{\pri -> \bar{p}}} 
\end{equation}

The number of anti-protons can be obtained by integrating this expression over time. For the rest of the paper we return to the index $p$ for the primary particles.

An integral, such as integrating over the expression above, is relevant for the summed production of gammas, neutrinos, or anti-protons, in this region, the interaction zone, and it is

\begin{equation}
\int N_{\INTE, p} \, d t\; = \; N_{p, 0} \, \tau_{\INTE, \star, p} \, \left( 1 \, -  \, \exp{\{- \, t/\tau_{\INTE, \star, p}\}} \right) \, .
\end{equation}

Any rules, how the primary particle spectrum of rigidity translates into the spectrum of the particle produced have to be obeyed, when using this integral, just as when using the Kardashev loss limit \cite{Kardashev62}, which steepens the spectrum by unity after the appropriate integral over space, or time, and may introduce a cutoff. For the case of anti-protons, in agreement with earlier authors Protheroe \cite{Protheroe81} has shown that the spectrum of anti-protons becomes asymptotically close to the primary spectrum of protons (Fig. 1 in \cite{Protheroe81}). Gaisser et al. (eq. 32 in \cite{Gaisser95}) have discussed in detail how the $\gamma$- and neutrino-spectra as a result of interactions come close to the energetic particle spectra. 

The time scale $\tau_{\INTE, \star, p}$ has two limits, as explained above. The contribution from interaction corresponds to the injection spectrum in the lower energy range, and to the steepened spectrum at the high energy range. For protons, pion production losses may replace spallation losses in this reasoning. Here we make the simplifying assumption, that all integrals over rigidity of the particles to obtain the gamma, neutrino or anti-proton spectrum can be taken as local in rigidity, so that the local spectrum of particles can be simply translated into the spectrum of the resulting particles. This all implies that the spectrum of gammas, neutrinos and anti-protons in the source region is determined by the original injection spectrum of the nuclei at lower energies, and at higher energies the injection spectrum is steepened by diffusion losses in the source region. This says that protons and anti-proton spectra naturally have the same spectrum in the source region at lower energies, and {\it a fortiori} also the same spectrum for an observer. At higher energies these two spectra will diverge, with anti-protons steeper by the diffusion time energy dependence in the source region. The transition energy between these two regimes depends on the source parameters. For the two-zone model it will be necessary to go through these arguments again, to verify what is happening there. There is further interaction in molecular clouds not far from the original sources, with the same spectrum; however, as shown in the detailed treatment of CR propagation in the ISM in \cite{deBoer17}, there is also interaction with the average spectrum far from the sources, then steepened by the Kolmogorov law. 

The particles escaping to the observer or bubble zone are given by

\begin{equation}
\frac{d \, N_{\INTE, p, \obs}}{d \, t} \; = \; \frac{N_p}{\tau_{\INTE, \DIF, p}} \,  , 
\end{equation}

\noindent which yields

\begin{equation}
N_{\INTE, p, \obs} \; = \; N_{p, 0} \, \frac{\tau_{\INTE, \star, p}}{\tau_{\INTE, \DIF, p}} \,  \left(1 \, - \, \exp{\{ - \, t/ \tau_{\INTE, \star, p} \}} \right) , 
\end{equation}

So here, adding a bubble  zone, implies, that $N_{\INTE, p, \obs} \; -> \; N_{\bub, p}$.

This converges for large times to the original spectrum, as expected, diminished by the spallation losses, as for large times the factor is

\begin{equation}
 \frac{\tau_{\INTE, \star, p}}{\tau_{\INTE, \DIF, p}} \,  = \, \frac{\tau_{\INTE, \spa, p}}{\tau_{\INTE, \spa, p} \, + \, \tau_{\INTE, \DIF, p}}, 
\end{equation}

\noindent which converges for large energies or small time scales to unity. In other words, for small time scales there is only spallation, so little change in the spectrum from the original injection \cite{Ferrari14}.

\subsection{AMS cosmic-ray iron nuclei, Fe}

A trial fit to the AMS spectrum can be made for, say, iron \cite{AMS21a,AMS21f}: The curve to fit is the spectral shape for Fe given by AMS after taking out the primary spectrum of $E^{-2.7}$. The equation used to fit, on first try, is:

\begin{equation}
N_{\INTE, p, \obs} \; = \; \frac{\tau_{\INTE, \star, p}}{\tau_{\INTE, \DIF, p}} \, N_{p, 0} \, . \, 
\end{equation}
 
From the definition of $\tau_{\INTE, \star, p}$  it follows

\begin{equation}
\frac{\tau_{\INTE, \star, p}}{\tau_{\INTE, \DIF, p}} \; = \; \left(1 \, + \, \frac{\tau_{\INTE, \DIF, p}}{\tau_{\INTE, \spa, p}} \right)^{-1} \; -> \, \frac{\tau_{\INTE, \spa, p}}{\tau_{\INTE, \DIF, p}} \, 
\end{equation}

for low energies, and so gives a positive spectrum, a term rising with energy.

We have shown above that along the path from the sources to us, through the ISM, the grammage is negligible, matching the recognition, that the environment of the exploding stars with enriched winds is sufficient to provide the grammage necessary \cite{Sina01}. Therefore we exclude this from the trial function.

It will be assumed that $\tau_{\INTE, \spa, p}$ is a constant \cite{Ferrari14}, and that $\tau_{\INTE, \DIF,p}$ is given by a power-law, with some possible choices explained above. So a trial can be attempted with

\begin{equation}
\frac{1}{1 \, + \, b \, R_2^{-\, \alpha}} \, 
\end{equation}

\noindent where $R_2$ is the rigidity in units of 100 GV, and $b$ and $\alpha$ are constants to be determined. $b$ is the ratio of the two time-scales, the diffusion time and the spallation time, at 100 GV, and $\alpha$ is the power-law exponent of the rigidity dependence of the diffusion time. This trial shows that the exponent $\alpha$ has to be much larger than unity, to match the depletion of the Fe spectrum below 100 GV.


However, any significant spallation production of Fe is not allowed, since there is no very abundant element above Fe, that could contribute.

The secondary to primary ratio Li/C (\cite{AMS18}) suggests $\alpha \, = \, - \, 1/3$ in the interaction zone; this trial suggests that the data require a second zone, the bubble zone, with a much steeper diffusion time slope.

This means that no version of a single zone diffusion and spallation can accommodate the data. However, double zone spallation and diffusion can explain this as a combination of diffusion time running as ${\tilde{R}}^{-\, 1/3}$ in the interaction zone, and as a steeper law ${\tilde{R}}^{-\, 5/3}$ in the bubble zone. There is bound to be some interaction already in the SN-shock loaded up with with CRs before the encounter with the interaction zone; since there the time-scales are independent of energy \cite{CR-I,CR-II,CR-III,CR-IV}, this would show up as secondaries showing at high energy the same spectrum as the primaries. With three zones or more this tendency of combining terms for each zone continues, and would allow yet steeper spectra, or other combinations of spectra. Since there is a small level of interaction outside the bubble, in the near environment and extremely little in the general interstellar medium, as shown above, all the way to the observer, the model proposed here is a simplification of the true situation, but can be adapted well to the data. The AMS data allow to test the relevance of adding inner or outer zones to fit the data. 

There is one caveat here, if some Co or Ni nuclei would initially have an abundance of the same order of magnitude as Fe at high energy: These nuclei would have to be destroyed almost completely, to explain the very low observed abundances (e.g. \cite{Wiebel98}). Such an extreme case appears to be unlikely at the present stage of our knowledge.

\subsection{Interaction zone: secondaries}

With the definition

\begin{equation}
\frac{1}{\tau_{\INTE, \star, p, s}} \; = \, \frac{1}{\tau_{\INTE, \star, p}} \, - \, \frac{1}{\tau_{\INTE, \star, s}} \, = \, \frac{1}{\tau_{\INTE, \spa, p, s}}
\end{equation}

the expression for secondaries can be rewritten as

\begin{equation}
\exp{\{- \, t/\tau_{\INTE, \star, s}\}} \, \frac{d }{d \, t} \, {\left(\, \exp{\{+ \, t/\tau_{\INTE, \star, s}\}} \,N_{\INTE, s} \, \right)} \, = \,  \frac{N_{p, 0}}{\tau_{\INTE, \spa, p}} \, \exp{\{ -t/\tau_{\INTE, \star, p}\}}  
\end{equation}

This yields

\begin{equation}
N_{\INTE, s} \; = \; N_{p, 0} \, \frac{\tau_{\INTE, \spa, p, s}}{\tau_{\INTE, \spa, p}}  \, \left( \exp{\{- \, t/\tau_{\INTE, \star, s}\}} - \exp{\{- \, t/\tau_{\INTE, \star, p}\}}  \right) \,  
\end{equation}

\noindent and correspondingly

\begin{equation}
N_{\INTE, s, \obs} \; = \; \frac{\tau_{\INTE, \spa, p, s}}{\tau_{\INTE, \spa, p}} \, \frac{N_{p, 0}}{\tau_{\INTE, \DIF, s}}  \, \left(\tau_{\INTE, \star, s} \, \left( 1 \, - \, exp{\{- \, t/\tau_{\INTE, \star, s}\}} \right)\, - \, \tau_{\INTE, \star, p} \left( 1 \, - \, exp{\{- \, t/\tau_{\INTE, \star, p}\}} \right) \right)\, 
\end{equation}

\noindent which for large times converges to

\begin{equation}
\frac{N_{\INTE, s, \obs}}{N_{\INTE, p, \obs}} \; = \; \frac{\tau_{\INTE, \star, p, s}}{\tau_{\INTE, \spa, p}} \,  \left( \frac{\tau_{\INTE, \star, s}}{\tau_{\INTE, \star, p}} \, - \, 1 \right) \, . 
\end{equation}

Here it is assumed that $\tau_{\INTE, \DIF, p} \, = \, \tau_{\INTE, \DIF, s} $.

In the limit that there is no secondary spallation,  this converges to the standard picture, that this ratio runs with the diffusion time, so for Kolmogorov $\sim \, \tilde{R}^{- \, 1/3}$. 

To fit the data it is necessary to allow, that even He, C, and O may have a small component which is a spallation product of Fe. Fe itself illustrates how a small effect of spallation can flatten the spectrum \cite{Wiebel98}, and conversely a small component of secondaries can steepen or otherwise modify the spectrum. Finally the primary spectrum is not certain. Alternatively, of course, it is possible that other well established CR components explain this excess, like from ISM-SNe, so white-dwarf-SNe (SNe Ia) or neutron-star-SNe. 

The main CR spectra given by AMS for He, C, and O illustrate \cite{CREAM10,AMS21e,CALET22} that there is a flatter component at higher energy just as proposed in Stanev et al. \cite{CR-IV}, The interplay between this flatter component and the small onset of spallation gain as well as spallation loss has to be determined with an iterative fit.

The observed ratio between clear secondaries such as Li, Be, B and dominant primaries such as He, C, O and Fe shows that the diffusion times are well approximated by a Kolmogorov spectrum of irregularities (see \cite{Biermann15}). 

The weakening of the spectra of He, C, O, and Fe at low energies/rigidities are consistent with strong spallation losses, by an order of magnitude at lower energies \cite{Sina01}. Using again a Kolmogorov spectrum constraints can be derived on any secondary contribution to He, C, and O from Fe spallation.

\section{Two zone model}


Above it has been seen that it is necessary to consider two zones, to allow for both the Li/C ratio, as well as the strong reduction of the primary fluxes below 100 GeV/$A$.  So next it is assumed that there is a second zone with spallation and diffusion, the bubble zone. The interaction zone lets particles diffuse into the bubble zone and, from there, particles diffuse into the observer's zone. From the observer's point of view this is injection. A key free parameter is that the diffusion-time energy dependence in the two zones could be different; in fact this is what the AMS observations suggest, as noted above.
 
So there the first additional equation is

\begin{equation}
\frac{d \, N_{\bub, p}}{ d \, t} \; = \; \frac{N_{\INTE, p}}{\tau_{\INTE, \DIF, p}} \, - \, \frac{N_{\bub, p}}{\tau_{\bub, \DIF, p}} \, - \, \frac{N_{\bub, p}}{\tau_{\bub, \spa , p}} \, 
\end{equation}

\noindent which can be written as

\begin{equation}
\frac{d \, N_{\bub, p}}{ d \, t} \; = \; \frac{N_{\INTE, p}}{\tau_{\INTE, \DIF, p}} \, - \, \frac{N_{\bub, p}}{\tau_{\bub, \star , p}}  . 
\end{equation}

This can be rewritten as

\begin{eqnarray}
& \frac{d}{d \, t} \, \left( \exp{\{{+t}/{\tau_{\bub, \star, p}}\}} \; N_{\bub, p} \right) \; = \; \frac{N_{p, 0}}{\tau_{\INTE, \DIF, p}} \, \exp{\{- \, {t}/{\tau_{\INTE, \star, p}} \, + {t}/{\tau_{\bub, \star, p}}\} }\; = \;\cr & \frac{N_{p, 0}}{\tau_{\INTE, \DIF, p}} \, \exp{\{-{t}/{\tau_{\INTE, \bub, \star, p}}\}} . 
\end{eqnarray}

with

\begin{equation}
1/{\tau_{\INTE, \bub, \star, p}} \, = \, 1/{\tau_{\INTE, \star, p}} \, - \, 1/{\tau_{\bub, \star, p}} \, .
\end{equation}

The solution can be written as

\begin{equation}
N_{\bub, p} \; = \; N_{p,0} \, \frac{\tau_{\INTE, \bub, \star, p}}{\tau_{\INTE, \DIF, p}} \, \left( \exp{\{- \, {t}/{\tau_{\bub, \star, p}}\}} \, - \, \exp{\{- \, {t}/{\tau_{\INTE, \star, p}}\}}\right) . 
\end{equation}

Here it is noted that a hierarchy of time scales is necessary, so the numbers always stay positive, so that  $\tau_{\bub, \star, p} \, > \, \tau_{\INTE, \star, p}$. 

The time-integral necessary to consider neutrino, gamma and anti-proton production is given in the long time limit by

\begin{equation}
\int \, N_{\bub, p} \, d \, t \; = \; N_{p,0} \, \frac{\tau_{\INTE, \bub, \star, p}}{\tau_{\INTE, \DIF, p}} \, \left({\tau_{\bub, \star, p}} \, - \, {\tau_{\INTE, \star, p}}\right) . 
\end{equation}

In the limit of $\tau_{\INTE, \star, p} \, >> \, \tau_{\INTE, \spa, p}$, and ${\tau_{\bub, \star, p}} \, >> \, {\tau_{\INTE, \star, p}}$ so typically at low energy,  the first term reverts to $\{\tau_{\INTE, \spa, p}\}/\{\tau_{\INTE, \DIF, p}\}$, whereas at high energy, so $\tau_{\INTE, \star, p} \, << \, \tau_{\bub, \spa, p}$  this ratio reverts to unity. In the same two limits of low and high energy the second term approaches the spallation time scale at low energy and the diffusion time scale at high energy. Combining these two limits implies that at low energy the energy dependence of this integral is the inverse diffusion time scale energy dependence, and at high energy the direct energy dependence of the diffusion time scale.

Then the particles diffusing to the outside are given by

\begin{equation}
\frac{d \, N_{\obs, \bub, p}}{d \, t} \; = \; \frac{N_{\bub, p}}{\tau_{\bub, \DIF, p}} . 
\end{equation}

which gives

\begin{equation}
N_{\obs, \bub, p} \; = \; N_{p, 0} \, \frac{\tau_{\INTE, \bub, \star, p}}{\tau_{\INTE, \DIF, p}} \, \left[ \frac{\tau_{\bub, \star, p}}{\tau_{\bub, \DIF, p}} \, \left(  1 - \exp{\{- \, {t}/{\tau_{\bub, \star, p}}\}} \right) \, - \, \frac{\tau_{\INTE, \star, p}}{\tau_{\bub, \DIF, p}} \,  \left(  1 \, - \, \exp{\{- \, {t}/{\tau_{\INTE, \star, p}}\}} \right)\right] . 
\end{equation}

Taking this to long times and allowing for the interaction zone time-scales to be much shorter than in the bubble-zone times gives

\begin{equation}
\frac{N_{\obs, \bub, p}}{N_{p, 0}} \; = \; \frac{\tau_{\INTE, \bub, \star, p}}{\tau_{\INTE, \DIF, p}} \, \frac{\tau_{\bub, \star, p}}{\tau_{\bub, \DIF, p}}
\end{equation}

The first fraction can (again) be written as with $\tau_{\bub} \, >> \tau_{\INTE}$

\begin{equation}
\frac{\tau_{\INTE, \star, p}}{\tau_{\INTE, \DIF, p}} \; = \; {\left({1 + \frac{\tau_{\INTE, \DIF, p}}{\tau_{\INTE, \spa, p}}}\right)}^{-1} \, ,
\end{equation}

and the second fraction as

\begin{equation}
\frac{\tau_{\bub, \star, p}}{\tau_{\bub, \DIF, p}} \; = \; {\left({1 + \frac{\tau_{\bub, \DIF, p}}{\tau_{\bub, \spa, p}}}\right)}^{-1} \, .
\end{equation}

Since the times in the bubble zone are typically much longer than in the interaction zones, the other two terms can be neglected. 
This means that with bubble zone similar factors can be gotten a second time, diffusion time over spallation time, so enhancing the terms which lead to inverted spectra at low energy.  We note that for any effect on the CR spectrum the ratio of time-scales is critical: so if the time scales in the bubble zone are all larger than in the interaction zone by a corresponding factor, then the effect flattening a CR spectrum can be the same for both zones, only different because of the different exponents of the rigidity dependence.   These spectra now can explain a steeper inversion. Again, the key ratios converge to unity at large energies, so that the source spectrum is reproduced, as expected.

For a fit this means that

\begin{equation}
\frac{N_{\obs, \bub, p}}{N_{p, 0}} \; = \; \; {\left(1 \, + \, b_{I} R_2^{- \, \alpha_I}\right)}^{-1} \, {\left(1 \, + \, b_{B} R_2^{- \, \alpha_B}\right)}^{-1}
\end{equation}

where

\begin{equation}
b_I \; = \; \frac{\tau_{\INTE, \DIF, p}(100 \, {\rm GV})}{\tau_{\INTE, \spa, p}(100 \, {\rm GV})} \;\;\; {\rm and} \;\;\; 
b_B \; = \; \frac{\tau_{\bub, \DIF, p}(100 \, {\rm GV})}{\tau_{\bub, \spa, p}(100 \, {\rm GV})} \, .
\end{equation}



It is to be noted that here we assume that the spallation times are independent of rigidity, a reasonably good assumption above a fraction of GeV \cite{Ferrari14}, who explicitly state,  that the total elastic and nonelastic cross sections for hadron-nucleus collisions are approximately constant (verbatim).

$\alpha_I$ and $\alpha_B$ are corresponding exponents, the rigidity dependence of the two diffusion times.

In the limit of $1 \,<< \, b_{I} R_2^{- \, \alpha_I}$ and $1 \,<< \, b_{B} R_2^{- \, \alpha_B}$ this implies that the overall expression runs as $R_2^{+ \, \alpha_I \, + \, \alpha_B}$, which can be quite steep.

 In case $b_I $ and  $b_B$ are small, then the relative deviations from unity at $R_2 \; = \, 1$ are $- \, \left(b_I \, + \, b_B\right)$. Below we show a simple fit, with a combination of Kolmogorov in the interaction zone (so $\alpha_I \, = \, 1/3$) and the lightning model (so $\alpha_B \, = \, 5/3$). 

For the bubble zone 

\begin{equation}
\frac{d \, N_{\bub, s}}{d \, t} \; = \; \frac{N_{\INTE, s}}{\tau_{\INTE, \DIF, s}} - \frac{N_{\bub, s}}{\tau_{\bub, \DIF, s}} \, + \, \frac{N_{\bub, p}}{\tau_{\bub, \spa, p}} - \frac{N_{\bub, s}}{\tau_{\bub, \spa, s}} .  
\end{equation}

Here furthermore:

\begin{equation}
1/\tau_{\INTE, \bub, \star, p} \, = \, 1/\tau_{\INTE, \star, p} \, - \, 1/\tau_{\bub, \star, p} \,\,\,{\rm and} \,\,\, 1/\tau_{\INTE, \bub, \star, p, s} \, = \, 1/\tau_{\INTE, \star, p} \, - \, 1/\tau_{\bub, \star, s} \, 
\end{equation}

\begin{equation}
1/\tau_{\INTE, \bub, \star, s} \, = \, 1/\tau_{\INTE, \star, s} \, - \, 1/\tau_{\bub, \star, s} \,\,\, {\rm and} \,\,\, 1/\tau_{\bub, \star, p, s} \, = \, 1/\tau_{\bub, \star, p} \, - \, 1/\tau_{\bub, \star, s} \, 
\end{equation}

and also

\begin{equation}
1/\tau_{\INTE, \star, p, s} \, = \, 1/\tau_{\INTE, \star, p} \, - \, 1/\tau_{\INTE, \star, s}
\end{equation}

and simplifiying

\begin{equation}
1/\tau_{\bub, \star, p, s} \, = \, 1/\tau_{\bub, \spa, p, s} \,\,\, {\rm and} \,\,\,
1/\tau_{\INTE, \star, p, s} \, = \, 1/\tau_{\INTE, \spa, p, s} \, 
\end{equation}

The last term here relates to spallation of one secondary particle population to another secondary particle population. The process of adding particles in the balance equation from the spallation of other secondary particle is ignored, since for such a term would have to distinguish two or more secondary particles.  This is obviously a very crude model, as an entire network of secondary products should be considered.

From this 

\begin{equation}
\exp{\{- \, t/\tau_{\bub, \star, s}\}} \, \left(\frac{d }{ d \, t} \, N_{\bub, s} \, \exp{\{+ \, t/\tau_{\bub, \star, s}\}}\right)\; = \; \frac{N_{\INTE, s}}{\tau_{\INTE, \DIF , s}} \, + \, \frac{N_{\bub, p}}{\tau_{\bub, \spa, p}}  \, 
\end{equation}

This evolves to 

\begin{eqnarray}
& \frac{d}{d \, t} \, \left( \exp{\{{+t}/{\tau_{\bub, \star, s}}\}} \; N_{\bub, s} \right) \; = \;  \frac{N_{p, 0}}{\tau_{\INTE, \DIF, s}} \, \frac{\tau_{\INTE, \spa, p, s}}{\tau_{\INTE, \spa, p}} \, \left(\exp{\{ - \, {t}/{\tau_{\INTE, \bub, \star, s}}\}} \,  - \, \exp{\{ - \, {t}/{\tau_{\INTE, \bub, \star, p, s}}\}} \right) \, \cr & + \, \frac{N_{p, 0}}{\tau_{\INTE, \DIF, p}} \, \frac{\tau_{\INTE, \bub, \star, p}}{\tau_{\bub, \spa, p}} \, \left(\exp{\{ - \, {t}/{\tau_{\bub, \spa, p, s}}\}} \, - \, \exp{\{ - \, {t}/{\tau_{\INTE, \bub, \star, p, s}}\}} \right) 
\end{eqnarray}

and finally to

\begin{eqnarray}
& & N_{\bub, s}  \; = \;  \frac{N_{p, 0}}{\tau_{\INTE, \DIF, s}} \, \frac{\tau_{\INTE, \spa, p, s}}{\tau_{\INTE, \spa, p}} \, {\tau_{\INTE, \bub, \star, s}} \, \left( \exp{\{ - \, {t}/{\tau_{\bub, \star, s}}\}} \,  - \, \exp{\{ - \, {t}/{\tau_{\INTE, \star, s}}\}} \right) \,  \cr  & - &
\frac{N_{p, 0}}{\tau_{\INTE, \DIF, s}} \, \frac{\tau_{\INTE, \spa, p, s}}{\tau_{\INTE, \spa, p}} \, {\tau_{\INTE, \bub, \star, p, s}} \,  \left( \exp{\{ - \, {t}/{\tau_{\bub, \star, s}}\}} \,  - \, \exp{\{ - \, {t}/{\tau_{\INTE, \star, p}}\}} \right)  \, \cr & - &\, \frac{N_{p, 0}}{\tau_{\INTE, \DIF, p}} \, \frac{\tau_{\INTE, \bub, \star, p}}{\tau_{\bub, \spa, p}} \,  {\tau_{bub, \spa, p, s}} \, \left(\exp{\{ - \, {t}/{\tau_{\bub, \star, s}}\}} \,  - \, \exp{\{ - \, {t}/{\tau_{\bub, \star, p}}\}} \right) \,  \cr & + &\, \frac{N_{p, 0}}{\tau_{\INTE, \DIF, p}} \, \frac{\tau_{\INTE, \bub, \star, p}}{\tau_{\bub, \spa, p}} \, {\tau_{\INTE, \bub, \spa, p, s}} \, \left(\exp{\{ - \, {t}/{\tau_{\bub, \star, s}}\}} \, - \, \exp{\{ - \, {t}/{\tau_{\INTE, \star, p}}\}} \right)  . 
\end{eqnarray}

To obtain the particles finally going outside $N_{\obs, \bub, s}$

\begin{equation}
\frac{d \ N_{\obs, \bub, s}}{d \, t} \; = \; \frac{N_{\bub, s}}{\tau_{\bub, \DIF, s}} \, 
\end{equation}

and then going to infinite times four terms are obtained multiplying $N_{p, 0}$, to get $N_{\obs, \bub, s}$:

Term 1:

\begin{equation}
\frac{\tau_{\INTE, \spa, p, s}}{\tau_{\INTE, \DIF, s}} \, \frac{\tau_{\bub, \star, s} \, - \, \tau_{\INTE, \star, s}}{\tau_{\bub, \DIF, s}} \, \frac{\tau_{\INTE, \bub, \star, s}}{\tau_{\INTE, \spa, p}} \, 
\end{equation}

Term 2:

\begin{equation}
\frac{\tau_{\INTE, \spa, p, s}}{\tau_{\INTE, \DIF, s}} \, \frac{\tau_{\INTE, \bub, \star, p, s}}{\tau_{\bub, \DIF, s}} \, \frac{\tau_{\INTE, \star, p} \, - \, \tau_{\bub, \star, s}}{\tau_{\INTE, \spa, p}} \, 
\end{equation}

After allowing that the time scales in the bubble-zone are all much longer than in the interaction zone, combining these two terms gives

\begin{equation}
\frac{\tau_{\bub, \spa, p, s}}{\tau_{\INTE, \DIF, s}} \, \frac{\tau_{\bub, \star, s}}{\tau_{\bub, \DIF, s}} \, \frac{\tau_{\INTE, \bub, \star, s}}{\tau_{\INTE, \spa, p}}  \, 
\end{equation}

This corresponds to the secondaries formed in the interaction zone, and transported through the bubble zone to the outside.

Term 3:

\begin{equation}
\frac{\tau_{\INTE, \bub, \star, p}}{\tau_{\INTE, \DIF, p}} \, \frac{\tau_{\bub, \star, s} \, - \, \tau_{\bub, \star, p}}{\tau_{\bub, \DIF, s}} \, \frac{\tau_{\INTE, \bub, \spa, p, s}}{\tau_{\bub, \spa, p}} \, 
\end{equation}

Term 4:

\begin{equation}
\frac{\tau_{\INTE, \bub, \star, p}}{\tau_{\INTE, \DIF, p}} \, \frac{\tau_{\INTE, \bub, \star, p, s}}{\tau_{\bub, \DIF, s}} \, \frac{\tau_{\INTE, \star, p} \, - \, \tau_{\bub, \star, s}}{\tau_{\bub, \spa, p}} \, 
\end{equation}

Combining these two terms in turn gives:

\begin{equation}
\frac{\tau_{\INTE, \bub, \star, p}}{\tau_{\INTE, \DIF, p}}   \, \frac{\tau_{\bub, \star, s}}{\tau_{\bub, \DIF, s}}\, \, \frac{\tau_{\INTE, \bub, \spa, p, s}}{\tau_{\bub, \spa, p}} 
\end{equation}

This corresponds to the secondaries formed in the bubble zone, and transported to the outside.

Note, that the primaries, as derived earlier, have the term

\begin{equation}
\frac{\tau_{\INTE, \star, p}}{\tau_{\INTE, \DIF, p}}  \, . 
\end{equation}

and $\tau_{\INTE, \bub, \star, p}$ is dominated by $\tau_{\INTE, \star, p}$, so the first terms are almost identical.

Comparing with the expression for the primaries it is found that the first two terms are reproduced in both expressions here, and that the third term for the interaction zone, $({\tau_{\INTE, \bub, \star, s}})/({\tau_{\INTE, \spa, p}})$ turns into $\tau_{\INTE, \DIF, s}/\tau_{\INTE, \spa, p}$ for high rigidities and using the assumption that all time scales in the bubble zone are much longer than in the interaction zone. This reproduces then the power-law akin to the Kolmogorov spectrum. The second term, $({\tau_{\bub, \star, s}})/({\tau_{\bub, \DIF, s}})$for the production of secondaries in the bubble zone the extra term devolves to $1/\left(1 \, + \, \tau_{\bub, \spa, p}/\tau_{\bub, \DIF, s}\right)$, again at high rigidity. This term runs as unity for small ratios of these two time-scales, and as the inverse of this ratio for large values; in our model, that the exponent of the diffusion time in the bubble zone runs as ${\tilde{R}}^{-5/3}$ and so implies, that these secondary particles rapidly become invisible relative to the secondaries from the interaction zone. This term gives a steeper spectrum for secondaries at low energies, as actually detected (discussed in \cite{ASR18}), but whether this fits quantitatively remains to be tested. The secondaries produced in the bubble zone may be below detectability, providing a serious constraint on the parameters.

These suggestions can be understood as follows: The interaction zone is given by the shocked region directly defined by the SN-shock hitting the wind-shock, a chemically heavily enriched region. Its turbulence may be dominated by the largest scale, giving a Kolmogorov spectrum. The outer zone can be interpreted as a wind-SN Superbubble \cite{Binns85,Binns89,Binns01,Binns05,Binns06,Binns07,Binns08,Binns11a,Binns11b,Binns13,FermiLAT11}, with turbulence mostly excited by lightning, the bubble zone.

\subsection{AMS cosmic-ray Li/C}

The best data to match is the published plot of the spectral indices \cite{AMS21g,Seo22}.

The secondary to primary ratio Li/C \cite{AMS18} suggests $\alpha \, = \,  1/3$ in a single zone; detailed data in fact suggest that at lower energy this ratio suggests a slightly steeper energy dependence $\alpha \, = \,  5/9$ \cite{ASR18}. That is the spectral index produced when a cosmic ray spectrum of $E^{- \, 7/3}$ by wave-particle-interaction gives a wave spectrum of $k^{-13/9}$ \cite{Biermann01}. If the secondaries from the bubble zone were to give this steepening, that index would be larger, then $\alpha \, = \,  1$, possibly already excluded by the data.

Following the analysis of CR Fe \cite{AMS21a,AMS21c,AMS21f} above suggests that reality needs at least two zones, in our proposal the interaction zone and the bubble zone. Can the ratio be reproduced with this energy dependence in that case?

Comparing the three ratios of time scales in the two terms for $N_{\obs, \bub, s}$ versus $N_{\obs, \bub, p}$, it can be noted that for $N_{\obs, \bub, p}$ the first ratio refers to first the interaction zone, and second to the outer shell, the bubble zone.  This confirms that the two zone model can reproduce the results of the single zone model at low energies for the CR Li/C ratio, using the Kolmogorov spectrum for the diffusion in the interaction zone. The Li nuclei produced in the bubble zone have a spectral behavior too steep to be visible under the spectrum of those produced in the interaction zone.

There is one difference at high energy: At high energy $\tau_{\bub, \star, s, s}$ approaches first $\tau_{\INTE, \star, s}$, and with it then $\tau_{\INTE, \spa, s}$, a constant, so this model predicts that the Li/C ratio should approach a constant in both models, in the single zone as well as in the two zone model. On the other hand, if $1/\tau_{\INTE, \spa, s}$ approaches zero, then this ratio keeps running as $\tau_{\INTE, \DIF, s}$ in both models.

The flatter component of the primaries (what is called the polar cap component in \cite{CR-I,CR-II,CR-III,CR-IV}) begins to dominate at higher energy (predicted to be $\sim \, E^{+ \, 1/3}$ flatter than at lower energy), so $E^{- \, 2}$ at source. Then for $Li/C$ running as $E^{- \, 1/3}$ implies that Li runs as $E^{- \, 8/3}$ in observations. This upturn is indeed very close to what is seen.

\subsection{Comparison with primaries}

This expression for secondaries has to be compared with the corresponding term for primaries $N_{\obs, \bub, p}$

\begin{equation}
\frac{\tau_{\bub, \star, p, p}}{\tau_{\INTE, \DIF, p}} \, \frac{\tau_{\bub, \star, p}}{\tau_{\bub, \DIF, p}} \, 
\end{equation}

It can be noted that there are only four time scales for a primary, and two independent time scales for a secondary, the diffusion and spallation time scales in each zone. All other time scales are composites. These time scales are assumed to obey following simple rules.

In the comparison note, that the power-law dependence of the diffusion time in the interaction zone (denoted with index $\INTE$)is determined and in the bubble zone (denoted with index $\bub$ ), which of course runs the same for primaries and secondaries in each of the two zones. This implies that there are only six independent time scales, three per zone. Also the time scales have a hierarchy, so that both time scales, for spallation and for diffusion, in the outer zone are slower, possibly by a similar factor. Finally it is assumed that spallation for secondaries  is slower than for primaries, but diffusion is the same. Therefore the ratio of the leading term for $N_{\obs, \bub, s}/N_{\obs, \bub, p}$ (e.g. the CR Li/O ratio) in the two zone model has to match what is derived in the one-zone model.

\section{Explaining the AMS cosmic-ray spectra and other cosmic-ray data}

First of all it is impressive, how the Fermi data (Fermi-Coll. 2016) as well as the ISS-CREAM spectrum \cite{ISS21a,ISS21b,ISS22} for CR protons make a clear and consistent prediction of what IceCube is expected to detect, under the assumption that the $\gamma$-ray emission observed is mostly hadronic. But the bump in the CR proton spectrum \cite{ISS21a,ISS21b,ISS22} at around 20 TeV just below the energies to which the IceCube data refer at their low energy boundary is another important aspect. A tentative proposal is that the CR-protons at really low energy are all secondary.  As was pointed out in \cite{ASR18},  their spectrum supports such an interpretation. The three source classes, ISM-SN-CRs, RSG-wind-SN-CRs, and BSG-wind-SN-CRs \cite{CR-IV,Biermann95} have suggested energy maxima that are quite different: ISM-SN-CRs are consistent with the maximum at around 50 TeV or slightly less, and the wind-SN-CRs both have a kink at 4 PeV, and a cutoff at 500 PeV (see \cite{Gaisser13,DAMPE19}). So in this approach the bump seen in the CR proton spectrum at around 20 TeV is interpreted as the maximum energy for ISM-SNe; those are both the SN Ia, or wd-SNe, and the neutron-star-SNe. That actually matches what was argued in \cite{CR-IV}, and all the higher energy protons are from wind-SNe. The population identified by the fit as "Pop 0" \cite{Gaisser13} has been fitted again in \cite{Bowman22}, and could be secondaries \cite{ASR18}. The difference between RSG-wind-SN-CRs and BSG-wind-SN-CRs is the chemical composition, with BSG-wind-SN-CRs heavily enriched. Their magnetic field is the same \cite{ASR18}, so their maximal energy $E_{max, p}$ should be the same. But of course since RSG-wind-SN-CRs may suffer a bit from adiabatic losses, their maximal energies might be scattered to slightly lower energies than for BSG-wind-SN-CRs, as was discussed in \cite{ASR18}. The detailed proposal by \cite{Gaisser13} can be well fitted to the data of CR protons and Helium nuclei.

Already, above, Fe and Li have been discussed, key tests for the proposed interpretation of the new AMS data.

\subsection{AMS cosmic-ray He, C, O}

If He, C, and O run with spallation losses just as Fe, then their spectra should be just shifted as their spallation cross section is smaller, but otherwise parallel. Clearly, that is not sufficient to explain the data, as there is extra curvature in the spectra.

The  comparison of CR O with CR Fe \cite{AMS21a,AMS21c,AMS21e,AMS21f} suggests that there is an extra component, also visible in CR He, and CR C. The source of this small extra component is not fully clear, but a comparison with CR protons suggests that there is no other component with such characteristics, arising from other CR sources, such as ISM SNe - either SN Ia, probably interacting white dwarfs, or those SNe, that make neutron stars. Therefore this leads to the idea that this could be a small spallation component, adding to these CR nuclei.

This concept is consistent with the observation by AMS \cite{AMS21c} that the $^3$He/$^4$He ratio has a rigidity dependence which is identical to that of B/O and B/C at high energy. The spectral index of $^3$He/$^4$He is found to be $= - \, 0.294 \, \pm \, 0.004$. This suggests here that this number is slightly modified from $- \, 1/3$ since $^4$He itself has a small secondary component with a somewhat different spectrum. If this secondary component in $^4$He, C, and O is within an order of magnitude similar to Li, Be, or B, it would slightly flatten the spectrum derived for the original $^4$He, C, and O CR spectrum, and increase the numerical value of the spectral index of the ratio $^3$He/$^4$He by a correspondingly small amount.

This means that the source spectrum for He, C, and O has to be corrected down by this small spallation component, making the original spectrum slightly flatter. There is no such component for Fe, and so a firm prediction of the model presented here is, that the corrected spectra for He, C, and O should match the Fe spectrum at energies at which spallation is certainly irrelevant. For this exercise to succeed an estimate is needed for the spallation loss of Fe, which then also modifies its spectrum.

\subsection{AMS cosmic-ray Ne, Mg, Si, and N, Na, and Al}

Prantzos has written extensively about some of these elements \cite{Prantzos86,Prantzos91,Prantzos93,Prantzos95,Prantzos11,Prantzos12a,Prantzos12b}. Here it is proposed that Ne, Mg and Si \cite{AMS21c,AMS21e} have a slightly larger secondary component mixed in than He, C, and O \cite{AMS20,AMS21d}. On the other hand N, Na and Al (\cite{AMS21c}) have a significant proportion of secondaries mixed in, as shown by the AMS fits.

\subsection{Anti-protons}


Above it was shown that the integral production of secondaries has two spectral regimes, with a spectrum at low energy which is the same as the primaries, and a steepened spectrum at high energy. This is consistent with both anti-proton spectra as well as the newly published neutrino spectra \cite{IceCube23}. It predicts that the anti-proton spectra will turn down at higher energy, and the neutrino spectra will turn flatter at lower energy. Again, a detailed treatment was performed in \cite{deBoer17} which included molecular clouds, and in this more general treatment CR interaction of the average CR spectrum in clouds gives rise to additional components in producing $\gamma$s, neutrinos, and anti-protons. \cite{deBoer17,Breuhaus22} required both interaction near the sources, the process at the center here, and also interaction of the average CR particles with molecular clouds.

\subsection{Gamma emission}

Fermi and other satellites have surveyed the gamma emission from the Galaxy, discussed by various authors, e.g. Fermi-LAT Coll. \cite{FermiLAT11,FermiLAT16,Boer17}. There the various emission contributions are listed \cite{Boer17} and compared with the model, including now both RSG and BSG stars, as well as other sources. A related model was proposed by \cite{Bykov22}.

First it should be noted that, of course, the penetration of molecular clouds from the average CR population will produce spectra consistent with their spectrum. That strong effect is not included in the considerations above. It matches the observed gamma spectrum  \cite{Boer17} and also predicts the neutrino spectrum correctly \cite{IceCube23}. The main interaction zone both in the one-zone model as well as the two zone model predict at lower energies a spectrum consistent with the particle injection spectrum (consistent with the relatively flat spectrum of the Cygnus wind-SN-Superbubble, \cite{FermiLAT11,FermiLAT23} and at higher energies a spectrum steepened by the diffusion time energy dependence, consistent with both Fermi-LAT and observed by IceCube \cite{IceCube23}. The Fermi-LAT data suggest that there is a small flatter component \cite{Boer17}, and that is in fact expected by the analysis above for the lower energies. The Galactic Center region offers analogous evidence for a flat spectral component, close to the expected source spectrum of CRs \cite{VERITAS21}.

In addition, the model presented here gives a turn-over of the spectrum at low energies, from the inverse of the diffusion time energy dependence to a flatter spectrum, to give at higher energy the direct diffusion energy dependence in addition to the source spectrum; the second part matches what is observed. The low energy behavior presents a break at low energy, visible in Fermi-LAT \cite{FermiLAT11} data of the Cygnus wind-SN Superbubble, and in Fig 4 of de Boer et al. \cite{Boer17} referred to as the MCR break, and the model presented here offers a different interpretation.

It is possible to work out estimates of the relevant time scales, for spallation, and diffusion, from the numbers given in Fermi-LAT \cite{FermiLAT11}, and they are quite reasonably short; however, the uncertainties are too large to provide a good test.

\subsection{Neutrinos}

Neutrino emission from sources in the Galaxy has been studied extensively already before \cite{Berezinsky93,Gaisser95,Becker08,Breuhaus22}, with the discovery in \cite{IceCube23}.

For protons, the spallation term can be re-interpreted as pion production loss, but when $\tau_{\mathit{pion}, p} \, >> \, \tau_{\INTE, \DIF, p}$, in the interaction zone the integral spectrum is steeper than injection, so that the ensuing neutrino emission is steeper, as well as the anti-proton spectrum. 

This interpretation predicts that, for the wind-SN-CR contribution, that there should be a kink in both the anti-protons as in the gammas and neutrinos, for the anti-protons towards a steeper spectrum at higher energies, and for gammas and neutrinos towards a flatter spectrum at lower energies. For the gammas this is well established.

\subsection{Electrons, positrons, and the First Ionization Potential effect}

Electrons and positrons are thoroughly discussed in \cite{ASR18}, with the extra positrons accounted for by triplet pair production, high energy electrons interacting with photons of sufficient energy \cite{Haug75,Haug81,Haug85,Haug04}. The production spectrum could be fitted to the data; considering models with extra losses, the observed spectrum may have to be slightly modified. The main parameters in such a fit are the maximal electron energy, the typical photon energy and the photon energy density; other weaker parameters are the electron spectrum and the photon spectrum. The numbers implied for the photon energy, its energy density and the maximum electron energy are quite reasonable. So there is a classical explanation for these positrons.

The effect of the first Ionization Potential is also discussed in \cite{ASR18}, and it is shown to be derivable from the selection effects of the plasma physics at injection.

\subsection{More zones?}

The question invariably arises whether two zones could be bested by three or even four or five zones. That becomes then akin to radiation transport in stellar atmospheres, when there is a transition to a treatment of the transport of energy. How would one recognize the necessity to use more zones? One key assumption is that there is no transport back down, so in a two-zone model there is all transport up from zone 1 to zone 2, and from zone 2 to the observer. The astronomical setting suggests that there could be a zone 3 in the form of a giant molecular cloud surrounding the wind-SN super bubble, a cloud zone. One could consider also the initial SN-shock which delivers the initial CR-distribution to the first zone, as another site where particles interact, and produce secondaries, the shock zone. One could treat the ISM traversed on the path from source to observer as a fourth zone, the ISM-zone. The approach adopted here is to see whether the data can be described by the simplest scenario possible. In this context, this suggests, that for explosions of RSG stars a one-zone model may be adequate; for a BSG star explosion the data suggest at least two zones, the shock shell interaction zone, and the wind-SN super bubble, the bubble zone (e.g. \cite{Binns05,Binns06,Binns07}).  The data do not require more than two zones yet at the level with the first simple fits. As long as the time-scales for all such zones obey a simple hierarchy, some of the expressions stack as multiple factors, all with the same basic structure. The fit to the Fe-CR-spectrum shown below requires that about half of the zones have a steep spectrum for scattering, as steep as given by lightning, see below. On the other hand, the direct data on secondaries suggest a Kolmogorov spectrum, which of course yields a flatter spectrum, which then dominates over a steeper spectrum over much of the rigidity range observed. Therefore two different zones are required. 

If the lightning model could be excluded, then more zones would be required to get the very steep decline of the primary spectra below 100 GV, because each zone would add a term like

\begin{equation}
{\left(1 \, + \, b_{I} R_2^{- \, \alpha_I}\right)}^{-1}
\end{equation}

Since the lightning model allows to fit the data, more zones are not required at this stage. For $b_{z} R_2^{- \, \alpha_z}\, >> \, 1$, where the index $z$ denotes any zone,  the two zones in combination give a term $R_2^{+ \, 2}$ for Kolmogorov plus lightning. To match that without lightning we would requite saturation twice and once Kolmogorov to make it sufficient; however, then the exponent in combination would give $7/3$, which would not fit the data anymore.

\subsection{A lightning model fit to the CR Fe spectrum}

A trial fit is shown in figure 1, using these two dependencies ${\tilde{R}}^{- \, 1/3}$ for the interaction zone, and ${\tilde{R}}^{- \, 5/3}$ for the bubble zone.

\begin{figure}[htpb]
\centering
\includegraphics[scale=0.45]
{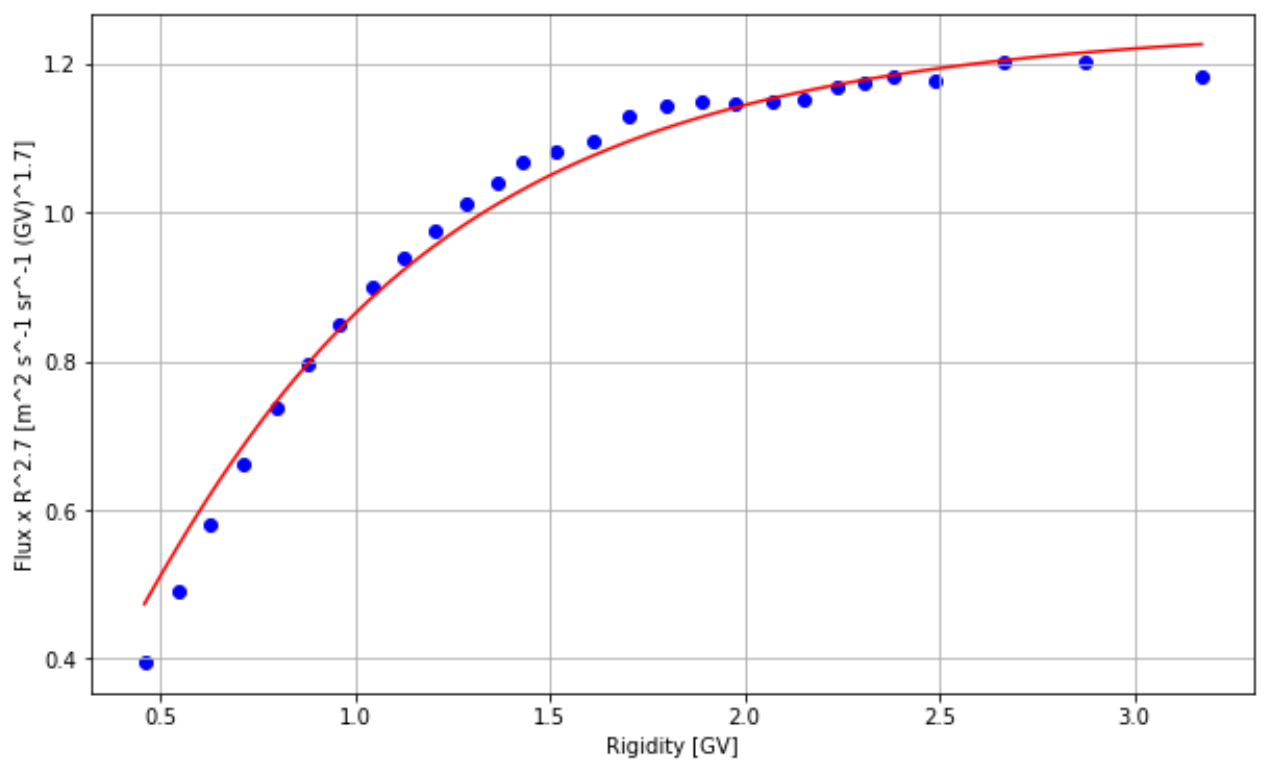}
\caption{A trial fit to the Fe spectrum $(1 + a R_2^{-1/3})^{- \, 1} \times (1 + b R_2^{-5/3})^{- \, 1}$ with $a \, = \, b \, = \, 0.03$, obtained from AMS \cite{AMS21a,AMS21f}, in flux multiplied with ${\tilde{R}}^{+ \, 2.70}$, so in units of ${\rm m^{-2} \, s^{-1} \, sr^{-1} (GV)^{1.70}}$ and the abscissa in rigidity in GV; both are in the decimal logarithm. This graph demonstrates that such a steep rigidity dependence of the diffusion in the bubble zone is required to fit the data, consistent with the lightning model: The low rigidity slope of the data in the graph approaches the steep slope of +2 of the ansatz.}
\end{figure}
\label{2024AthinaMelifitAMSF}

 This fit has only two parameters, the absolute normalization, and the parameter $b$, the ratio of the diffusion time scale to the spallation time scale at a rigidity of 100 GV in each of the interaction zone, and in the bubble zone, assumed for the purpose of this demonstration to be the same number. Because of the uncertainties in how many BSG star contribute to the CR flux no error bars are shown; if that number were modest, then there could be wiggles in the spectrum unaccounted for in the model.

This ansatz is capable of matching the steep decline in CR flux below 100 GV, relative to the injected spectrum. It needs to be emphasized that the rigidity where the two terms, $b_I \, {\tilde{R}}^{- \, 1/3}$ and $b_B \, {\tilde{R}}^{- \, 5/3}$, become equal to unity, so where this term becomes less relevant at higher values of the rigidity, could in principle be at very different rigidities. This trial has been done ignoring the upturn \cite{CR-I,CR-II,CR-III,CR-IV}, visible in the AMS data for CR H, CR C, and CR O \cite{AMS21e}.

\section{Conclusion}

A model of CR interaction has been presented in one and two source zones, the interaction zone and the bubble zone. In this time-dependent model the spectra as a function of time are worked out, and integrated to obtain the emission over time or go towards infinite time for the particle population released. This approach permits a more detailed set of spectra, which allow interpretation and tests, such as the Fermi-LAT data \cite{FermiLAT11,FermiLAT16,FermiLAT23,VERITAS21} discussed in de Boer et al. \cite{Boer17}. This is consistent with the very high fraction of core-collapse SNe coming from OB-Superbubbles (e.g. \cite{Higdon98,Higdon05,Binns05,Binns06,Binns07,Lingenfelter07,Binns08,Higdon13}).

It is obvious that the layer cake model should be improved quantitatively by using the known spallation and interaction cross-sections, for all elements and their isotopes. Furthermore, currently there is no stage of adiabatic expansion or adiabatic compression, that would give a shift down or up in rigidity. The rather naive approximation of pion or meson production in proton-proton collisions with a time-scale to remove a proton considered from its specific bin in rigidity analogously to spallation can be improved. A full approach taking all such refinements into account would be best treated with a large Monte-Carlo code, with every bin of rigidity for every isotope treated separately. 

Comparing with the AMS data fitting the primary CR elements He to Fe requires two source zones: The inner zone, the interaction zone, needs a diffusion regime, which appears to follow the Kolmogorov law, and the outer zone, the bubble zone, which is consistent with the spectrum of magnetic irregularities excited by lightning; this lightning arises from the variable jets/winds of freshly formed BHs within the OB-superbubble. Lightning produces a characteristic steep spectrum of particles \cite{Gopal24} and a corresponding flat spectrum of magnetic irregularities. This then also predicts the spectrum of the primary elements, as well as the secondary CR elements. The model implies that He, C, and O CR elements have a small secondary component; this is qualitatively supported by the detection of secondary $^3$He, obeying the same rules as CR element B. Here the sequence from He, C, and O to Ne, Mg, Si, and further to N, Na, and Al has been proposed as CR elements progressively including more secondaries. The test would be to show that the enriched matter column necessary for producing these secondaries of each element is actually the same; of course it could happen, that we have to distinguish several different types of BSG stars; this is now possible \cite{Chieffi13,Limongi18,Limongi20}. These elements derive from Blue Super Giant (BSG) star explosions. This sequence of more spallation contributions is the main prediction of the model, and it can be falsified by showing otherwise, for instance if the column required were very different for each element beyond using several mass ranges for BSG stars \cite{Chieffi13,Limongi18,Limongi20}, would point to implausible numerical values, or would just fail to fit the data.

Red Super Giant (RSG) star explosions may provide a strong contribution to the anti-protons, gammas and neutrinos.  In this model the anti-protons are expected to turn down in their spectra relative to protons at higher energy. Neutrinos in turn are expected to turn to a flatter spectrum at lower energy. 

To completely test the model, the following steps are necessary:

1) Test the two-zone model approach using Fe, and determine the diffusion time rigidity exponents to see, whether ${\tilde{R}}^{- \, 1/3}$ and ${\tilde{R}}^{- \, 5/3}$ are really the best fit.

2) If so, there is only one free parameter per element left to determine all the other spectra, since the spallation cross sections are known, and the ratio between the interaction zone and the bubble zone is given by Fe. So then a global fit, matching all elements with the same parameters, should be possible. The source abundances derived, in turn, should match, what is expected from stellar evolution nucleosynthesis simulations \cite{Chieffi13,Limongi18,Limongi20}.

3) Work out the secondary spectra as a further test. All columns are known from the data fits of the earlier steps here. These two steps, (2) and (3), interact, so these two steps may require an iterative fit. All elements should show both gains and losses except for the heaviest nuclei, for which we expect only losses.

4) Work out the contributions to $\gamma$s, neutrinos, and anti-protons, which should never exceed what is seen. As we noted here are additional contributions from CR interaction in clouds.

5) Check the normalizations for all elements and isotopes to see whether the simulations by \cite{Chieffi13} are sufficient. It is possible that other sources also contribute. For protons and He that is already obvious, as they have components from ISM-SNe, as noted above.

6) Work out the polar cap components for all elements and isotopes to predict what higher energy data should detect, in CR spectra, $\gamma$s, neutrinos and anti-protons.

7) Extend the modeling to EeV energies to test it with Auger data, such as \cite{Auger20}.

8) Test, whether the CR population dubbed pop 3 by \cite{Gaisser13} can be detected in the data, and whether its anti-proton component can be constrained. That is a component predicted by the lightning model. This component appears to arise from the neighborhood of rotating black holes.

9) Check whether the model can be extended to electrons and positrons, beyond what has been done in \cite{ASR18}.

10) Develop an analogous model for starburst galaxies such as M82 to test whether the observations can yield further constraints on the models. If yes, then apply the model to the early phase in galaxy evolution, when starbursts were common in the early universe.

Afterwards, if the model turns out to give a reasonable approximation and understanding of the data, then one might consider the next steps:

Work through the interaction beyond the bubble-zone, and model the ISM passage including all molecular clouds, and at the same time remain consistent with the Fermi data \cite{deBoer17}. Reintegrate the model to deal with starburst galaxies, both at the current epoch like M82, and in the early universe.

However, there are caveats and pitfalls, that may make this series of steps more difficult and even impossible:

i) Our concept of just two zones, the interaction zone and the bubble zone, with one spectrum of magnetic irregularities each, must surely be far too simple. Often, there are molecular clouds nearby to OB-super-bubbles.

ii) Our concept of just using RSG and BSG star explosions of single stars, perhaps of several different flavors, is surely far too simple. Such stars are almost all in stellar binary systems, triples or quadruples \cite{Chini12,Chini13a,Chini13b}, an aspect ignored here.

iii) What is the contribution of those SNe, that produce neutron stars, which in turn can accelerate particles visibly in pulsar wind nebulae? These stelar explosions contribute visibly to CR protons and CR He, as noted.

iv) The most common type of SN just included here as ISM-SNe here are SN Ia; what do they contribute beyond CR protons and some CR He?

v) The pathways going into molecular clouds and back out \cite{Appenzeller74,deBoer17} give rise to $\gamma$-ray emission that is detected. So how much do they contribute in CR interaction? How much do they contribute to the neutrinos detected?  

vi) Micro-quasars \cite{Mirabel99} quite visibly contribute energetic particles: In fact their jets imply electric currents, which are driven by an energetic particle spectrum, of spectrum $E^{-2}$ in the model proposed by \cite{Gopal24}. These particles are ejected, and may have been seen in the Auger data \cite{Auger20}. So much more do micro-quasars contribute?

vi) CRs are also accelerated by AGN, even when quiescent \cite{Chini89,Saikia18}, just only weakly visible. Quiescent AGN encompasses all massive galaxies. How much is this contribution?

viii) Our Galaxy is believed to have a wind (see, e.g., for observable consequences like the WMAP haze \cite{Biermann10}) and surely that has a strong influence on energetic particles coming in from outside, and distributing some energetic particles from Galactic sources all over the disk.  Furthermore, that wind is not steady. How much can that energetic particle contribution be? How can such an unsteady wind contribute in CR propagation?

ix) The star formation in our Galaxy is very patchy both in space and in time (see, e.g., \cite{Smith78}). How much does that skew the population of energetic particles visible today at Earth? 

x) Our Galaxy does a relatively modest mass super-massive black hole (SMBH) at its center, which to all appearances never had a merger merger with any other SMBH. Activity of our SMBH is very erratic. Can past activity episodes have contributed to what we observe today in energetic particles?

How can any proposal to account for energetic particle properties possibly be so simple?

The proposal is a first step and predicts contributions from one kind of massive star explosions, the RSG and BSG star explosions. ISM-SNe are necessary to account for one population of CR protons and CR He.



\section{Acknowledgements}

This paper is dedicated to the memories of Tom Gaisser. The origin of this paper can be traced back to discussions in 1992 and 1993 between Tom Gaisser, Todor Stanev and Peter Biermann, leading to publications near that time. Since the 1970s to recently inspiring discussions between Phil Kronberg and Peter Biermann kept on pushing the limits of what these radio super\-novae in the starburst galaxy M82, discovered in 1981,  can teach us. Peter Biermann wishes to thank Nasser Barghouty, Julia Becker Tjus, Silke Britzen, Roland Diehl, Gopal-Krishna, Matthias Kaminski, Wolfgang Kundt, Norma Sanchez, Gary Webb, and many others for discussions.  The authors thank Roger Clay and two referees for incredibly helpful detailed comments on the manuscript.


\section{References}

\end{document}